\documentclass[prx,aps,superscriptaddress,footinbib,twocolumn]{revtex4-2}

\usepackage{mathrsfs}
\usepackage{amsfonts}
\usepackage{amssymb}
\usepackage{amsmath}
\usepackage{amsthm}
\usepackage{graphicx}
\usepackage[usenames,dvipsnames]{color}
\usepackage[colorlinks=true,citecolor=blue,linkcolor=magenta]{hyperref}
\usepackage{lmodern}
\usepackage{dsfont}
\usepackage{natbib}
\usepackage{bm}
\usepackage{tabularx}
\usepackage{multirow}
\usepackage{tabularray}
\usepackage{comment}
\usepackage{bbm}
\usepackage{mathtools}

\usepackage{tikz} 
\usetikzlibrary{quantikz2}

\usepackage{algcompatible}
\usepackage{algorithm}

\algrenewcommand\algorithmicindent{2em}

\theoremstyle{definition}

\theoremstyle{remark}

\renewcommand{\Re}{\mathrm{Re}}
\renewcommand{\Im}{\mathrm{Im}}
\newcommand{\Tr}{\mathrm{Tr}}
\newcommand{\norm}[1]{\Vert #1 \Vert}
\newcommand{\normLR}[1]{\left\Vert #1 \right\Vert}
\newcommand{\abs}[1]{\vert #1 \vert}

\newcommand{\ketbra}[2]{\vert #1 \rangle \langle #2 \vert}
\newcommand{\mean}[1]{\langle #1 \rangle}

\usepackage{soul}

\renewcommand{\arraystretch}{1.5}
\renewcommand{\leq}{\leqslant}
\renewcommand{\geq}{\geqslant}

\DeclareUnicodeCharacter{2009}{\,}

\begin{document}
	
	\title{Nonorthogonal variational quantum simulation for quantum chemistry}

	\author{Zongkang Zhang}
	\affiliation{Hefei National Research Center for Physical Sciences at the Microscale and School of Physical Sciences, University of Science and Technology of China, Hefei 230026, China}
	\affiliation{Shanghai Research Center for Quantum Science and CAS Center for Excellence in Quantum Information and Quantum Physics, University of Science and Technology of China, Shanghai 201315, China}
	\affiliation{Hefei National Laboratory, University of Science and Technology of China, Hefei 230088, China}
	
	\author{Jiajun Ren}
	\affiliation{Key Laboratory of Theoretical and
    Computational Photochemistry, Ministry of Education, College of
    Chemistry, Beijing Normal University, Beijing 100875, China}
	
	\author{Xiao Yuan}
	\email{xiaoyuan@pku.edu.cn}
	\affiliation{Center on Frontiers of Computing Studies, School of Computer Science, Peking University, Beijing 100871, China}

	\begin{abstract}
		Dynamical simulation of quantum many-body systems is a central task in quantum chemistry and requires efficient wavefunction representations. Although tensor-network and neural-network quantum states have achieved considerable success in ground-state calculations, entanglement growth hinders their application to quantum dynamics. Quantum computing may offer a route to quantum advantage, but algorithms such as Trotterization generally require fault-tolerant quantum computers, whereas near-term hardware supports only limited circuit sizes. Variational quantum simulation (VQS) instead represents the evolving state with a \emph{single} shallow, fixed-size parameterized quantum circuit (PQC). Here, we introduce nonorthogonal variational quantum simulation (NOVQS), which applies linear combinations of parameterized quantum states to real- and imaginary-time evolutions. We design a shallow, hardware-friendly ansatz tailored to second-quantized electronic-structure Hamiltonians, together with resource-efficient protocols for measuring the matrices and vectors in the parameter equations of motion. Error analysis and resource estimation are also provided. Numerical simulations of hydrogen chains and the nitrogen molecule demonstrate that a collection of shallow, or even single-layer, PQCs can match or outperform a much deeper PQC in VQS. In particular, we identify a trade-off between circuit number and depth. NOVQS therefore provides a flexible route to enhancing wavefunction expressivity under circuit-depth constraints, making it a promising framework for quantum dynamics simulations on near-term quantum processors.
	\end{abstract}

	\maketitle

	\section{Introduction}
	Accurately simulating quantum many-body dynamics governed by the Schr\"odinger equation is of fundamental importance in chemistry and materials science~\cite{Georgescu2014Quantum,McArdle2020Quantum}. The exponential complexity of wavefunction severely limits the scalability of conventional methods to large systems, with the rapid growth of entanglement during time evolution further exacerbating this computational challenge. A middle ground is to make full use of limited computational resources to efficiently characterize the states and dynamics of interacting quantum systems, thereby enabling their behavior to be accurately elucidated.
	
	In this direction, quantum Monte Carlo (QMC) methods can perform real- or imaginary-time simulations with polynomial computational complexity via stochastic sampling~\cite{Austin2011Quantum,becca2017quantum}. Based on their treatment of the wavefunction, QMC can be broadly classified into two categories. The first category, including diffusion Monte Carlo and auxiliary-field quantum Monte Carlo, employs a fixed trial wavefunction and uses an ensemble of random walkers to simulate the stochastic dynamics. However, for generic fermionic and frustrated systems, these methods suffer from the sign problem, which is NP-hard~\cite{Troyer2005computational}. The second category, referred to as time-dependent variational Monte Carlo (tVMC), constructs a parameterized ansatz for the wavefunction and evolves its time-dependent parameters according to the variational principle. Because the probabilities used to sample configurations are non-negative, tVMC is free from the sign problem. The ansatz provides an efficient compressed representation of the quantum state, and its expressive power largely determines the performance of tVMC. 
	
	In recent years, two families of ans\"atze have developed rapidly: tensor network states (TNS)~\cite{White1992Density,Ulrich2011,Chan2011DMRG,Orus2019} and neural quantum states (NQS)~\cite{Giuseppe2017Solving,Pfau2020,Hermann2023Ab,Lange2024From}. TNS can efficiently simulate the dynamics of one-dimensional systems~\cite{Vidal2004,Haegeman2011Time}, but face challenges arising from entanglement when applied to higher-dimensional systems~\cite{Eisert2015}. By leveraging the extensive toolbox of machine learning, NQS have emerged as a powerful framework for simulating quantum many-body systems~\cite{Hermann2023Ab,Lange2024From}. Despite considerable success in ground-state calculations, the use of NQS for simulating quantum dynamics is still under active investigation. In addition, evidence suggests that TNS and NQS may require an exponentially large number of parameters to simulate time evolution and to solve ground-state problems~\cite{Lin2021Real,Passetti2023Can}.
	
	In parallel, quantum computing offers a promising route towards fundamentally addressing quantum many-body problems~\cite{Feynman1982}. The inherent unitarity of quantum circuits makes them naturally suited to quantum simulation. Trotter–Suzuki product formulas are widely used for quantum simulation, but its circuit depth generally increases with the evolution time, the number of Hamiltonian terms, and the target accuracy~\cite{Seth1996,Ollitrault2021}. Quantum circuits of this scale would require universal fault-tolerant quantum computers and are therefore beyond the capabilities of near-term quantum hardware.
	
	Variational quantum simulation~\cite{Li2017efficient,McArdle2019Variational,Yuan2019theoryofvariational,Endo2020variational,Ollitrault2021,Gomes2021Adaptive,Lee2022Simulating,Lee2022Variational,Dobrautz2024Toward,Wan2024Hybrid,Huang2025Towards,Li2025Multiset} provides a hybrid quantum-classical framework that is particularly well suited to quantum computers with limited circuit depth. It utilizes a parameterized quantum circuit (PQC) as the ansatz wavefunction and updates the parameters by a classical computer. Both VQS and tVMC realizes the real- or imaginary-time evolution by optimizing time-dependent parameters according to a variational principle. Their principal distinction lies in how the variational state is represented and evaluated: tVMC typically employs classically tractable ans\"atze, such as TNS or NQS, and evaluates observables through Monte Carlo sampling, whereas VQS uses a PQC and performs the required measurements on quantum computers. In this sense, variational simulation can serve as a platform for benchmarking the expressive capabilities of quantum and classical networks.
	
	As a variational quantum algorithm~\cite{Cerezo2021}, the performance of VQS relies critically on the structure of PQCs. First, a natural strategy is to construct problem-specific ansatz. For example, the unitary coupled-cluster ansatz employed in the variational quantum eigensolver (VQE) is specifically tailored to molecular ground-state calculations~\cite{Wecker2015Progress}. VQS commonly adopts the Hamiltonian ansatz~\cite{Li2017efficient,Yuan2019theoryofvariational,Endo2020variational,Huang2025Towards}, as its circuit structure mirrors that of Trotterization. However, a molecular electronic Hamiltonian generally contains $\mathcal{O}(N_q^4)$ terms~\cite{McArdle2020Quantum}, where $N_q$ is the number of spin orbitals, potentially resulting in a prohibitively large circuit depth even for a single layer of the Hamiltonian ansatz. Designing ans\"atze suitable for quantum chemistry simulations therefore is an important problem. Second, and more importantly, the size of currently available quantum circuits is limited. Once an individual quantum circuit has been optimized to its fullest extent, is there still a way to further enhance the expressive power of the ansatz? In fact, a similar idea has already been explored in the context of VQE, where the target quantum state is represented as a linear combination of multiple nonorthogonal quantum states. This approach is known as the nonorthogonal variational quantum eigensolver (NOVQE)~\cite{Huggins2020,Baek2023say,Ren2026Error}. How such a nonorthogonal multistate ansatz can be extended to VQS, however, remains an open question.
	
\begin{figure*}[htbp]
		\centering
		\includegraphics[width=1.0\linewidth]{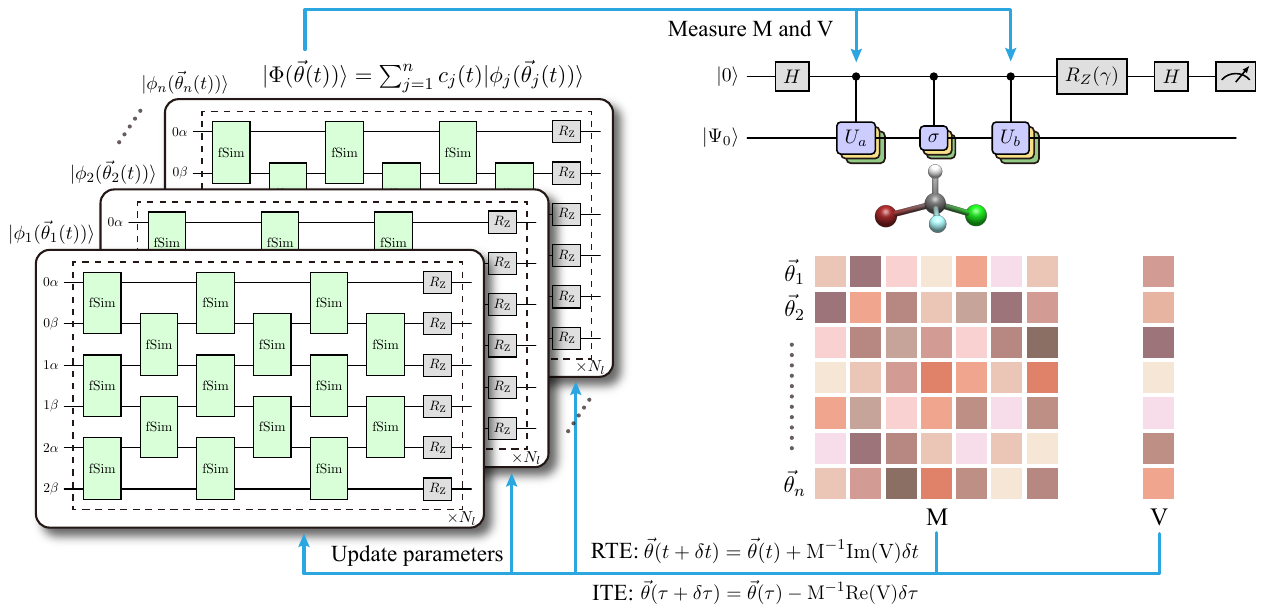}
		\caption{Schematic of nonorthogonal variational quantum simulation (NOVQS). The algorithm proceeds as follows: First, the variational wavefunction is represented by a set of parameterized quantum circuits constructed from fermionic simulation (fSim) gates. Second, the quantum Fisher information matrix $M$ and the force vector $V$ are evaluated on quantum computers. Subsequently, the parameters at time $t+\delta t$ are updated from their values at time $t$ using $M$ and $V$, applicable to both real and imaginary time evolution ($\tau \gets t$). This iterative process is repeated for $N_t = T/\delta t$ steps, where $T$ denotes the total evolution duration.}
		\label{fig:schematic}
	\end{figure*}

	In this work, we propose a nonorthogonal variational quantum simulation (NOVQS) framework for quantum chemistry, as illustrated in Fig.~\ref{fig:schematic}. Based on the variational principle, we derive the equations of motion (EOM) for NOVQS, which govern the time-dependent parameters in both real- and imaginary-time evolution. Inspired by Ref.~\cite{fSim2018Kivlichan}, we design a compact ansatz tailored to the simulation of electronic structure. We further develop improved measurement schemes: in addition to the conventional Hadamard test, the required quantities can be evaluated using ancilla-free or classical-shadow-based approaches, thereby removing the ancilla qubit and reducing the circuit depth and making NOVQS more suitable for near-term quantum hardware. Then, we provide error analysis and resource estimation for NOVQS. Finally, we perform numerical simulations of real- and imaginary-time simulation for molecular systems of different sizes. By comparing NOVQS with conventional VQS, we demonstrate that the expressive power of the ansatz can be systematically enhanced by increasing the number of PQCs. For the same aggregate circuit depth, NOVQS employing multiple shallow or even single-layer PQCs can match or even outperform conventional VQS employing a single much deeper PQC. In particular, we find that NOVQS exhibits a trade-off between the number of PQCs and their circuit depth.

	\section{Variational principle for unnormalized wavefunctions}
	In quantum chemistry, the electronic structure Hamiltonian is typically expressed in the second-quantized formalism as~\cite{McArdle2020Quantum}
	\begin{align}
		H = \sum_{p,q} h_{p,q} a_p^\dag a_q  + \frac{1}{2} \sum_{p,q,r,s} h_{p,q,r,s} a_p^\dag a_q^\dag a_r a_s,
	\end{align}
	where $a_j^\dag$ ($a_j$) denotes the fermionic creation (annihilation) operator for the $j$-th spin-orbit. The first term accounts for the kinetic energy of the electrons and their Coulombic attraction to the nuclei, while the second term represents the electron-electron Coulomb repulsion. The factor of $1/2$ is introduced to avoid double-counting, consistent with the symmetry $h_{p,q,r,s} = h_{q,p,s,r}$. In quantum computing, such second-quantized Hamiltonian is mapped to a Pauli form Hamiltonian $H = \sum_{\nu=1}^{N_h} h_{\nu} P_{\nu}$, where $h_{\nu}$ are real coefficients and $P_{\nu}$ are Pauli product strings.
	
	Given the Hamiltonian, the Schr\"odinger’s equation 
	\begin{align}\label{eq:Schrodinger}
		\frac{d \ket{\Psi(t)}}{d t} = - i H \ket{\Psi(t)}
	\end{align}
	describes the dynamics of quantum many-body systems. The time evolved quantum state is
	\begin{align}
		\ket{\Psi(t)} = e^{-i H t} \ket{\Psi_0}.
	\end{align}
	The Trotterization approach decomposes the time evolution operator as the Lie-Trotter-Suzuki product formula
	\begin{align}
		e^{-iHt} \simeq \prod_{i=1}^{N_t} \left( \prod_{j=1}^{N_h} e^{-i h_{\nu} P_{\nu} \Delta t} \right).
	\end{align}
	However, the depth of the corresponding quantum circuit scales linearly with the number of Hamiltonian terms $N_h$ and the total evolution steps $N_t$. The linear combination of unitaries method expresses the time evolution operator in the summation form
	\begin{align}
		e^{-iH \Delta t} = \sum_j c_j U_j.
	\end{align}
	At each time step, only one unitary $U_j$ is sampled, which reduces the circuit depth significantly. However, the required sampling cost scales exponentially with the total steps $N_t$. 
	The variational quantum simulation (VQS) circumvents the above problems by representing the time evolved quantum state with a parameterized quantum state, i.e.
	\begin{align}
		\ket{\Psi(t)} \simeq \ket{\Phi(\theta(t))},
	\end{align}
	where $\ket{\Psi(\theta(t))}$ is prepared by a parameterized quantum circuit (PQC), i.e. ansatz, and $\theta(t)$ are time-dependent circuit parameters. Here, the evolution of the quantum state is mapped to the motion of the parameters in the quantum circuit. The VQS is very suitable for the near term quantum computers due to the low overhead of quantum resources. 
	
	The main challenge of VQS is to design a PQC that has efficient expressive power within limited circuit size. Such a challenge also arises in the variational quantum eigensolver (VQE), where substantial efforts have been devoted to developing physically motivated and adaptive ansatz for ground-state problems. Besides focusing on a single ansatz, there are methods extending the ansatz space through the linear combination of several states~\cite{Long2006General,Long2008Duality,Childs2012hamiltonian,Berry2015simulating}, e.g. quantum subspace expansion~\cite{McClean2017Hybrid,Colless2018Computation}, nonorthogonal quantum eigensolver~\cite{Huggins2020,Baek2023say,Ren2026Error} and variational quantum circuit Monte Carlo~\cite{Yang2024Maximizing}. Inspired by their success in VQE, we extend the ansatz of VQS to the subspace spanned by a collection of PQCs
	\begin{align}\label{eq:ansatz}
		\ket{\Phi(\theta(t))} = \sum_{j=1}^{N_c} c_j(t) \ket{\phi_j(\theta_j(t))},
	\end{align}
	where $N_c$ is the number of quantum circuits.
	Generally, $c_j(t)$ is a complex coefficient and $\ket{\phi_j(\theta_j(t))} = U_j(\theta_j(t)) \ket{\Psi_0}$, where $\ket{\Psi_0}$ is the initial state. Note that the component states in the linear combination are, in general, pairwise nonorthogonal. Because this wavefunction is unnormalized, we have
	\begin{align}\label{eq:ansatz_normalized}
		\ket{\Psi(t)} \simeq \ket{\tilde{\Phi}(\theta(t))} \coloneqq \frac{\ket{\Phi(\theta(t))}}{\norm{\ket{\Phi(\theta(t))}}}.    
	\end{align}
	We numerically demonstrate that an ensemble of several shallow circuits can outperform a single deep circuit within the VQS framework. By optimizing the variational parameters $\theta_j(t)$ and the expansion coefficients $c_j(t)$, one can accurately track the trajectory of the state's time evolution. In Sec.~\ref{sec:variational principle_rte} and \ref{sec:variational principle_ite}, the equations of motion for both real- and imaginary-time evolution are derived from McLachlan’s variational principle~\cite{McLachlan1964}.

	\subsection{McLachlan’s variational principle}\label{sec:variational principle_rte}
	The McLachlan’s variational principle aims to minimize the norm between the left hand side and the right hand side of Eq.~(\ref{eq:Schrodinger}). Supposing the state $\ket{\Psi(\theta(t))}$ is normalized, it gives
	\begin{align}
		\mathcal{L}=&\normLR{\left( \frac{d}{d t} + i H \right) \ket{\Psi(\theta(t))}}^2 \notag \\ 
		=&\normLR{\left( \sum_k \dot{\theta}_k \partial_k + i H \right) \ket{\Psi(\theta(t))}}^2 \notag\\
		=&\sum_k \dot{\theta}_k \left[ \sum_q \dot{\theta}_q \braket{\partial_k \Psi}{\partial_q \Psi} + i \bra{\partial_k \Psi} H \ket{\partial_q \Psi} + \textrm{c.c.} \right],
	\end{align}
	where we ignore the terms independent with $\dot{\theta}_k$. Let $\partial \mathcal{L} / \partial \dot{\theta}_k = 0$, we got the EOM for the parameters as
	\begin{align}\label{eq:RTE-motion-normalized}
		\sum_q \dot{\theta}_q \Re\braket{\partial_k \Psi}{\partial_q \Psi}  = \Im\bra{\partial_k \Psi} H \ket{\Psi}.      
	\end{align}

	According to Ref.~\cite{Correcting2025Gentinetta}, we can use two virtual parameters $\theta_0^R$ and $\theta_0^I$ to encode the normalized factor and the global phase. Then the total wave function is 
	\begin{align}
		\ket{\Psi(t)} = e^{\theta_0^R + i \theta_0^I} \ket{\Psi(\theta(t))}.
	\end{align}
	The partial derivative of the wave function with respect to the virtual parameters are
	\begin{align}
		\ket{\partial_{\theta_0^R} \Psi} =& \ket{\Psi} \notag \\
		\ket{\partial_{\theta_0^I} \Psi} =& i\ket{\Psi} 
	\end{align}
	When $k=0$, according to Eq.~(\ref{eq:RTE-motion-normalized}),
	\begin{align}
		\dot{\theta}_0^R \braket{\Psi}{\Psi} + \sum_{q>0} \dot{\theta}_q \Re\braket{\Psi}{\partial_q \Psi} =& 0 \label{eq:virtual-Re}\\
		\dot{\theta}_0^I \braket{\Psi}{\Psi} + \sum_{q>0} \dot{\theta}_q \Im\braket{\Psi}{\partial_q \Psi} =& -\bra{\Psi} H \ket{\Psi} \label{eq:virtual-Im}
	\end{align}
	For $k>0$, using Eq.~(\ref{eq:virtual-Re}) and (\ref{eq:virtual-Im}), the left hand side of Eq.~(\ref{eq:RTE-motion-normalized}) is 
	\begin{align}
		&\sum_q \dot{\theta}_q \Re\braket{\partial_k \Psi}{\partial_q \Psi} \notag\\
		=& \dot{\theta}_0^R \Re\braket{\partial_k \Psi}{\Psi} - \dot{\theta}_0^I \Im\braket{\partial_k \Psi}{\Psi} + \sum_{q>0} \dot{\theta}_q \Re\braket{\partial_k \Psi}{\partial_q \Psi} \notag\\
		=& -\sum_{q>0} \dot{\theta}_q \frac{\Re\braket{\Psi}{\partial_q \Psi} \Re\braket{\partial_k \Psi}{\Psi}}{\braket{\Psi}{\Psi}} \notag\\
		&+ \sum_{q>0} \dot{\theta}_q \frac{\Im\braket{\Psi}{\partial_q \Psi} \Im\braket{\partial_k \Psi}{\Psi}}{\braket{\Psi}{\Psi}} \notag\\
		& + \frac{\bra{\Psi}H\ket{\Psi} \Im\braket{\partial_k \Psi}{\Psi}}{\braket{\Psi}{\Psi}} + \sum_{q>0} \dot{\theta}_q \Re\braket{\partial_k \Psi}{\partial_q \Psi} \notag\\
		=& \sum_{q>0} \dot{\theta}_q \left( \Re\braket{\partial_k \Psi}{\partial_q \Psi} - \frac{\Re(\braket{\Psi}{\partial_q \Psi}\braket{\partial_k \Psi}{\Psi})}{\braket{\Psi}{\Psi}} \right) \notag\\
		&+ \frac{\bra{\Psi}H\ket{\Psi} \Im\braket{\partial_k \Psi}{\Psi}}{\braket{\Psi}{\Psi}}. 
	\end{align}
	Finally, the EOM for parameters in unnormalized wave function is
	\begin{align}\label{eq:RTE-motion-unnormalized}
		\sum_{q>0} \dot{\theta}_q M_{kq}^R= V_k^I,
	\end{align}
	where 
	\begin{align}
		M_{kq}^R =& \Re\left( \frac{\braket{\partial_k \Psi}{\partial_q \Psi}}{\braket{\Psi}{\Psi}} - \frac{\braket{\Psi}{\partial_q \Psi}\braket{\partial_k \Psi}{\Psi}}{\braket{\Psi}{\Psi}^2}  \right), \label{eq:rte-M}\\
		V_k^I =& \Im\left( \frac{\bra{\partial_k \Psi} H \ket{\Psi}}{\braket{\Psi}{\Psi}} - \frac{\bra{\Psi}H\ket{\Psi} \braket{\partial_k \Psi}{\Psi}}{\braket{\Psi}{\Psi}^2}\right). \label{eq:rte-V}
	\end{align}
	Here, $M$ is a positive-definite covariance matrix and often called quantum Fisher information matrix, and $V$ is a force vector. For real-time evolution, we optimize the parameters in the variational wavefunction following Eq.~(\ref{eq:RTE-motion-unnormalized}).
	
	\subsection{Imaginary time evolution}\label{sec:variational principle_ite}
	Through a Wick rotation of the time coordinate, $t=-i\tau$, the real time evolution (RTE) Eq.~(\ref{eq:Schrodinger}) is transformed into imaginary time evolution (ITE)
	\begin{align}\label{eq:ITE}
		\frac{d \ket{\Psi(t)}}{d \tau} = -H \ket{\Psi(t)},
	\end{align}
	which is also very important in quantum many-body problems. For varational imaginary time simulation~\cite{McArdle2019Variational}, the EOM is
	\begin{align}\label{eq:ITE-motion-normalized}
		\sum_q \dot{\theta}_q \Re\braket{\partial_k \Psi}{\partial_q \Psi}  = -\Re\bra{\partial_k \Psi} H \ket{\Psi}.      
	\end{align}
	Similarly, the NOVQS updates the variational parameters following
	\begin{align}\label{eq:ITE-motion-unnormalized}
		\sum_{q>0} \dot{\theta}_q M_{kq}^R= -V_k^R,
	\end{align}
	where 
	\begin{align}
		M_{kq}^R =& \Re\left( \frac{\braket{\partial_k \Psi}{\partial_q \Psi}}{\braket{\Psi}{\Psi}} - \frac{\braket{\Psi}{\partial_q \Psi}\braket{\partial_k \Psi}{\Psi}}{\braket{\Psi}{\Psi}^2}  \right), \label{eq:ite-M}\\
		V_k^R =& \Re\left( \frac{\bra{\partial_k \Psi} H \ket{\Psi}}{\braket{\Psi}{\Psi}} - \frac{\bra{\Psi}H\ket{\Psi} \braket{\partial_k \Psi}{\Psi}}{\braket{\Psi}{\Psi}^2}\right). \label{eq:ite-V}
	\end{align}
	For imaginary-time evolution, the variational parameters are optimized according to Eq.~(\ref{eq:ITE-motion-unnormalized}).
	In the context of variational Monte Carlo (VMC), this EOM corresponds to the stochastic reconfiguration~\cite{Sorella1998Green} method, which is widely used to find the ground state.
	Similarly, Eq.~(\ref{eq:RTE-motion-unnormalized}) aligns with time-dependent VMC (tVMC) for simulating real-time quantum dynamics.
	In this regard, NOVQS can be viewed as the quantum computing counterpart to classical (t)VMC.
	The primary distinction lies in the way of evaluating $M$ and $V$: (t)VMC estimates them by Monte Carlo sampling, while NOVQS leverages quantum computers to measure them directly. 
	Recent work has identified support-mismatch bias in estimators arising from the sampling distribution used in (t)VMC and introduced blurred sampling to address this issue~\cite{Wan2026Removing}.

	\section{Quantum circuit ansatz and measurement protocols}
	
	In this section, we introduce the PQC tailored for the variational quantum chemistry simulations. Then, we proposed three methods to evaluate the quantum Fisher information matrix and the force vector: the standard Hadamard test, the ancilla-free Hadamard test, and the classical shadow procedure.

	\subsection{Ansatz for chemical systems}
	When considering spin systems, the variational Hamiltonian ansatz is a natural choice for the quantum many-body dynamics. The variational Hamiltonian ansatz can be written as 
	\begin{align}
		\ket{\phi(\theta(t))} = \prod_{l=1}^{N_l} \prod_{\nu=1}^{N_h} e^{-i P_{\nu} \theta_{l,\nu}} \ket{\Psi_0},
	\end{align}
	where $N_l$ is the number of circuit layers, the parameters set $\theta=\{ \theta_{l,\nu}\}_{l=1, \nu=1}^{N_l, N_h}$ and are initialized to zeros. The number of terms in the Hamiltonian for spin models scales from $\mathcal{O}(N_q)$ to $\mathcal{O}(N_q^2)$. Furthermore, by partitioning these terms into high-density commuting groups, the required number of parameters can be kept extremely small. In contrast, the electronic structure Hamiltonians in quantum chemistry typically contain $\mathcal{O}(N_q^4)$ terms, which necessitates the search for a more suitable ansatz.
	
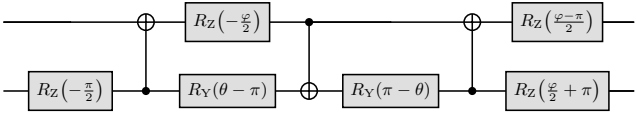
\begin{figure}[htbp]
		\resizebox{\columnwidth}{!}{
			\begin{quantikz}[row sep=0.55cm, column sep=0.45cm]
				& \qw
				& \targ{}
				& \gate[style={fill=gray!25}]{R_{\rm Z}\!\left(-\frac{\varphi}{2}\right)}
				& \ctrl{1}
				& \qw
				& \targ{}
				& \gate[style={fill=gray!25}]{R_{\rm Z}\!\left(\frac{\varphi-\pi}{2}\right)}
				& \qw
				\\
				& \gate[style={fill=gray!25}]{R_{\rm Z}\!\left(-\frac{\pi}{2}\right)}
				& \ctrl{-1}
				& \gate[style={fill=gray!25}]{R_{\rm Y}\!\left(\theta-\pi\right)}
				& \targ{}
				& \gate[style={fill=gray!25}]{R_{\rm Y}\!\left(\pi-\theta\right)}
				& \ctrl{-1}
				& \gate[style={fill=gray!25}]{R_{\rm Z}\!\left(\frac{\varphi}{2}+\pi\right)}
				& \qw
			\end{quantikz}
		}
		\caption{Quantum circuit decomposition of the fSim gate defined in Eq.~(\ref{eq:fSim}). The circuit implements the gate up to an irrelevant global phase $e^{-i\varphi/4}$, where $R_{\rm Z}(\alpha) = e^{-i\alpha Z/2}$ and $R_{\rm Y}(\alpha) = e^{-i\alpha Y/2}$.}
		\label{fig:fSim}
	\end{figure}

\begin{figure}[htbp]
		\centering
		\resizebox{\columnwidth}{!}{
			\begin{quantikz}[background color=green!15]
				\lstick{$0\alpha$} & \gate[2]{\text{fSim}} &                       & \gate[2]{\text{fSim}} &                       & \gate[2]{\text{fSim}} &                       & \gate[style={fill=gray!25}]{R_{\rm Z}} & \\
				\lstick{$0\beta$}  &                       & \gate[2]{\text{fSim}} &                       & \gate[2]{\text{fSim}} &                       & \gate[2]{\text{fSim}} & \gate[style={fill=gray!25}]{R_{\rm Z}} & \\
				\lstick{$1\alpha$} & \gate[2]{\text{fSim}} &                       & \gate[2]{\text{fSim}} &                       & \gate[2]{\text{fSim}} &                       & \gate[style={fill=gray!25}]{R_{\rm Z}} & \\
				\lstick{$1\beta$}  &                       & \gate[2]{\text{fSim}} &                       & \gate[2]{\text{fSim}} &                       & \gate[2]{\text{fSim}} & \gate[style={fill=gray!25}]{R_{\rm Z}} & \\
				\lstick{$2\alpha$} & \gate[2]{\text{fSim}} &                       & \gate[2]{\text{fSim}} &                       & \gate[2]{\text{fSim}} &                       & \gate[style={fill=gray!25}]{R_{\rm Z}} & \\
				\lstick{$2\beta$}  &                       &                       &                       &                       &                       &                       & \gate[style={fill=gray!25}]{R_{\rm Z}} &    
			\end{quantikz}
		}
		\caption{Structure of a single layer in the parameterized quantum circuit, taking $M=3$ spatial orbitals as an example. The ${\rm fSim}$ gates are arranged in a 'brickwork' pattern across $2M$ sublayers, followed by a final sublayer of $R_{\rm Z}$ rotation gates. Upon completion of one block, every pair of qubits has interacted exactly once, and the qubit ordering is reversed. The ${\rm fSim}$ gate is defined in Eq.~(\ref{eq:fSim}), and rotation gate $R_{\rm Z}(\lambda) = e^{-i\lambda Z/2}$.}
		\label{fig:ansatz_block}
	\end{figure}
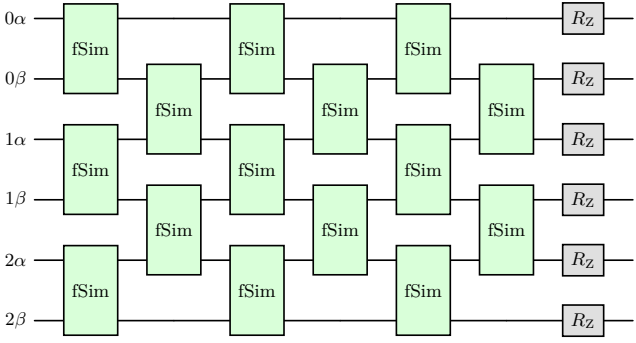

	To describe the dynamics of chemical system, we propose a dense local ansatz, which is inspired from the simulation of Fermionic Hamiltonian. The fermionic simulation gate, denoted as
	\begin{align}\label{eq:fSim}
		{\rm fSim}(\theta, \varphi) &= {\rm SWAP} \cdot {\rm CPhase(\varphi)} \cdot R_{\rm XX+YY}(-\theta) \notag\\  
		&=
		\begin{pmatrix}
			1 & 0 & 0 & 0 \\
			0 & i\sin{\theta} & \cos{\theta}   & 0 \\
			0 & \cos{\theta}  & i\sin{\theta}  & 0 \\
			0 & 0 & 0 & e^{i \varphi}
		\end{pmatrix},
	\end{align}
	is well suitable for electronic structure Hamiltonian simulation, e.g. Fermi-Hubbard model~\cite{fSim2018Kivlichan}. Here, $R_{\rm XX+YY}(-\theta) = e^{i\theta(X\otimes X+Y\otimes Y)/2}$, ${\rm CPhase}(\varphi)=e^{i\varphi (I-Z)\otimes(I-Z)/4}$ is the controlled phase gate, and SWAP gate swaps the states of two qubits. Experimental implementations of the fSim gate have been demonstrated using both superconducting qubits~\cite{Foxen2020Demonstrating,Morvan2022} and semiconductor spin qubits~\cite{Ni2025Diverse,tsoukalas2025resonant}. As a two qubit operation, fSim gate can be decomposed into at most $3$ entangling gates, such as CNOT or CZ gates, with some single qubit rotation gates~\cite{Vatan2004Optimal}; see Fig.~\ref{fig:fSim} for a detailed circuit decomposition. Through a dense brickwork pattern up to a layer of single qubit rotations, the circuit depth for a single Trotter step is only $N_q+1$ with $N_q(N_q-1)/2$ ${\rm fSim}$ gates, where $N_q$ is the number of spin orbitals and is even. For chemical molecules, the second-quantized Hamiltonian can be convert to a Hubbard-type Hamiltonian using double factorization representation~\cite{Cohn2021quantum,Oumarou2024accelerating}. Such transformation is based on the orbital rotation, which can be decomposed into Givens rotations~\cite{fSim2018Kivlichan,Cohn2021quantum}. Importantly, the compiled circuit for orbital rotation is similar to that of the fermionic Trotter step mentioned above. Therefore, we employ a PQC inspired from Ref.~\cite{fSim2018Kivlichan} as the ansatz, which is shown in Fig.~\ref{fig:ansatz_block}. This ansatz can be written as
	\begin{align}
		\ket{\phi(\theta, \varphi, \lambda)} =& 
		\prod_{l=1}^{N_l}\Bigg\{
		\left( \prod_{j=1}^{N_q} e^{-i \frac{\lambda_j}{2} Z_j} \right) \notag\\
		&\times \prod_{k=1}^{Nq/2} \Bigg[ 
		\left( \prod_{j\in S_{\rm odd}} {\rm fSim}_{j, j+1}(\theta_{k,j}, \varphi_{k,j}) \right) \notag\\
		&\times \left( \prod_{j\in S_{\rm even}} {\rm fSim}_{j, j+1}(\theta_{k,j}, \varphi_{k,j}) \right)
		\Bigg] \Bigg\} \ket{\Psi_0},
	\end{align}
	where $S_{\rm even}=\{0, 2, \cdots, N_q-2\}$ and $S_{\rm odd}=\{1, 3, \cdots, N_q-3\}$. Note that the qubit ordering of the final state is reversed whenever the number of layers $N_l$ is odd. This ansatz enables interactions between arbitrary pairs of spin orbitals with only linear circuit depth and connectivity, allowing efficient simulations with shallow circuits. It is therefore particularly well suited to near-term quantum processors with limited circuit depth and native connectivity restricted to nearest-neighbor interactions on a two-dimensional lattice.
	
	In general, the $j$-th PQC can be represented as an ordered sequence of gates
	\begin{align}
		U_j(\theta_j) = G_{j, N_{p,j}}(\theta_{j, N_{p,j}}) \cdots G_{j,k}(\theta_{j,k}) \cdots G_{j,1}(\theta_{j,1})
	\end{align}
	where $G_{j, k}(\theta_{j,k})$ represents the $k$-th gate with parameter $\theta_{j,k}$, and $N_{p,j}$ is the number of parameterized gates for the $j$-th ansatz. As a unitary operator, $G_{j,k}(\theta_{j,k})$ can be expressed as $e^{-i H \theta_{j,k}}$, where $H$ is a Hermitian operator and can be decomposed in the Pauli operator basis. Its derivative with respect to its parameter $\theta_{j,k}$ is
	\begin{align}\label{eq:partial_op}
		\frac{\partial G_{j,k}(\theta_{j,k})}{\partial \theta_{j,k}} = \sum_i f_{j,k,i} G_{j,k} \sigma_{j,k,i},
	\end{align}
	where $\sigma_{j,k,i}$ are Pauli product operators and $f_{j,k,i}$ are complex coefficients. Therefore, the tangent state with respect to the parameter $\theta_{j,k}$ is
	\begin{align}\label{eq:tangent}
		\frac{\partial \ket{\Phi(\theta)}}{\partial \theta_{j,k}} = \sum_i f_{j,k,i} R_{j,k,i} \ket{\Psi_0},
	\end{align}
	where the unitary $R_{j,k,i} = G_{j,N_{p,j}} \cdots G_{j,k} \sigma_{j,k,i} G_{j,k-1} \cdots G_{j,1}$. Note that the tangent state only relies on the $j$-th ansatz and thus the corresponding circuit also has $\mathcal{O}(N_{p,j})$ gates. 
	
	Our parameterized wavefunction consists of five categories of parameters: $\theta$ and $\varphi$ within the fSim gates, $\lambda$ for the $R_Z$ gates, as well as the amplitude and phase parameters associated with the complex coefficients $c$. Following Eq.~(\ref{eq:partial_op}), the partial derivative of the $R_Z(\lambda)$ gate is $\partial R_Z(\lambda) / \partial \lambda = -\frac{i}{2} R_Z(\theta) Z$. For the fSim gate, the gradients with respect to $\theta$ and $\varphi$ are given by: 
	\begin{align}
		\frac{\partial {\rm fSim}(\theta,\varphi)}{\partial \theta} &= \frac{i}{2} {\rm fSim}(\theta,\varphi) ({\rm X \otimes X} + {\rm Y \otimes Y}), \label{eq:partial_theta} \\
		\frac{\partial {\rm fSim}(\theta,\varphi)}{\partial \varphi} &= \frac{i}{4} {\rm fSim}(\theta,\varphi) ({\rm I\otimes I} - {\rm I \otimes Z} - {\rm Z \otimes I} + {\rm Z \otimes Z}). \label{eq:partial_phi}
	\end{align}
	Note that these expressions are derived considering that the ${\rm CPhase}$ and $R_{\rm XX+YY}$ operators commute.
	In general, $c_j(t)$ is a complex coefficient. We can parameterize it as $c_j(t)=a_j(t) e^{i b_j(t)}$. Then, the partial derivatives of the wavefunction with respect to $a$ and $b$ are respectively:
	\begin{align}
		\partial \ket{\Phi} / \partial  a_j &= e^{i b_j} \ket{\phi_j}, \label{eq:partial_a}\\
		\partial \ket{\Phi} / \partial  b_j &= i c_j \ket{\phi_j}. \label{eq:partial_b}
	\end{align}
	The total number of parameters and coefficients is given by $N_p = \sum_{j=1}^{N_c} N_{p,j} + 2 N_c $. By adopting an identical architecture for every PQC, this simplifies to $N_p = N_c (N_{p,j} + 2)$. For a circuit with $N_l$ layers, the number of parameters per PQC is $N_{p,j} = N_q^2 N_l$.

	\subsection{Estimation of $M$ and $V$}
	\subsubsection{Hadamard test}
	The construction of $M$ and $V$ requires measuring $\braket{\partial_k \Psi}{\partial_q \Psi}$, $\braket{\partial_k \Psi}{\Psi}$, $\bra{\partial_k \Psi} H \ket{\Psi}$, $\bra{\Psi}H\ket{\Psi}$, and $\braket{\Psi}{\Psi}$. These quantities can be efficiently estimated on a quantum computer via the Hadamard test~\cite{Li2017efficient,Ekert2002direct}.
	
	If the parameters labelled $k$ and $q$ are inside $\ket{\phi_j}$ and $\ket{\phi_{j'}}$ respectively, then we can write
	\begin{align}
		\braket{\partial_k \Psi}{\partial_q \Psi} &= \sum_{i,i'} f_{j,k,i}^* f_{j',q,i'} \bra{\Psi_0} R_{j,k,i}^\dag R_{j',q,i'} \ket{\Psi_0}, \label{eq:term1}	\\
		\braket{\partial_k \Psi}{\Psi} &= \sum_{i,j'} f_{j,k,i}^* c_{j'} \bra{\Psi_0} R_{j,k,i}^\dag U_{j'} \ket{\Psi_0}, \label{eq:term2}\\
		\bra{\partial_k \Psi} H \ket{\Psi} &= \sum_{i,\nu,j'} f^*_{j,k,i} h_{\nu} c_{j'} \bra{\Psi_0} R_{j,k,i}^\dag P_{\nu} U_{j'} \ket{\Psi_0}, \label{eq:term3} \\
		\bra{\Psi}H\ket{\Psi} &= \sum_{j,{\nu},j'} c_j^* h_{\nu} c_{j'} \bra{\Psi_0} U_j^\dag P_{\nu} U_{j'} \ket{\Psi_0}, \label{eq:term4} \\
		\braket{\Psi}{\Psi} &= \sum_{j,j'} c_j^* c_{j'} \bra{\Psi_0 }U_j^\dag U_{j'} \ket{\Psi_0}. \label{eq:term5} 
	\end{align}
	Each term in the summation can be expressed as $a e^{i\gamma} \bra{\Psi_0} U \ket{\Psi_0}$, where $a$ is a positive scalar and the remaining factor can be measured using Hadamard test.
	A schematic of the quantum circuits for measuring each term in Eqs.~(\ref{eq:term1})-(\ref{eq:term5}) are illustrated in Fig.~\ref{fig:hadamard_test}. The circuit concludes with a measurement of the ancilla qubit in the computational (i.e. Pauli-$Z$) basis. Each individual shot produces an outcome $\mu = \pm 1$, from which the final value is estimated as the statistical expectation $\mathbb{E}[\mu]$. By leveraging the identity $\Im (e^{i\gamma}\bra{\Psi_0}U\ket{\Psi_0}) = \Re (e^{i(\gamma - \pi/2)}\bra{\Psi_0}U\ket{\Psi_0})$, the imaginary part of each term can be straightforwardly measured by shifting the $R_{\rm Z}$ gate rotation angle to $\gamma - \pi/2$. In the case $k$ or $q$ indicates $a_j$ (or $b_j$), Eq.~(\ref{eq:tangent}) is replaced by Eq.~(\ref{eq:partial_a}) (or Eq.~(\ref{eq:partial_b})) and the unitaries $R_{j,k,i}$ are simply replaced by $U_j$.
	
	\begin{figure}[htbp]
		\centering
		
		\resizebox{\columnwidth}{!}{
			\begin{quantikz}
				\lstick{\ket{0}} & \gate{H} & \ctrl{1} & \ctrl{1} & \ctrl{1} & \gate{R_Z(\gamma)} & \gate{H} & \meter{} \\
				\lstick{\ket{\Psi_0}} & \qw & \gate{U_a} & \gate{\sigma} & \gate{U_b} & \qw & \qw & \qw
			\end{quantikz}
		}
		\vspace{1.5em}
		
		\setlength{\tabcolsep}{18pt} 
		\renewcommand{\arraystretch}{1.8} 
		\resizebox{\columnwidth}{!}{
			\begin{tabular}{cccc}
				\hline\hline 
				Expression                                  & $U_a$         & $\sigma$  & $U_b$            \\ \hline 
				$\braket{\partial_k \Psi}{\partial_q \Psi}$ & $R_{j',q,i'}$ & N/A       & $R_{j,k,i}^\dag$ \\ 
				$\braket{\partial_k \Psi}{\Psi}$            & $U_{j'}$      & N/A       & $R_{j,k,i}^\dag$ \\ 
				$\bra{\partial_k \Psi} H \ket{\Psi}$        & $U_{j'}$      & $P_{\nu}$ & $R_{j,k,i}^\dag$ \\
				$\bra{\Psi}H\ket{\Psi}$                     & $U_{j'}$      & $P_{\nu}$ & $U_j^\dag$       \\  
				$\braket{\Psi}{\Psi}$                       & $U_{j'}$      & N/A       & $U_j^\dag$       \\ \hline\hline
			\end{tabular}
		}
		\caption{Quantum circuit for the estimation of expressions in $M$ and $V$. The corresponding $U_1$, $\sigma$ and $U_2$ for Eqs.~(\ref{eq:term1})-(\ref{eq:term5}) are listed below. N/A means that no controlled-$\sigma$ gate is applied in this case. To measure the imaginary part, just change the rotation angle of the $R_{\rm Z}$ gate from $\gamma$ to $\gamma - \pi/2$.}
		\label{fig:hadamard_test}
	\end{figure}

	To estimate $\braket{\partial_k \Psi}{\partial_q \Psi}$, the number of required circuits scales as $\mathcal{O}(N_p^2 N_d^2)$, where $N_d$ denotes the number of terms in Eq.~(\ref{eq:tangent}). Similarly, the circuit requirements for $\braket{\partial_k \Psi}{\Psi}$, $\bra{\partial_k \Psi} H \ket{\Psi}$, $\bra{\Psi}H\ket{\Psi}$ and $\braket{\Psi}{\Psi}$ are $\mathcal{O}(N_c N_p N_d)$, $\mathcal{O}(N_c N_h N_p N_d)$, $\mathcal{O}(N_c^2 N_h)$ and $\mathcal{O}(N_c^2)$, respectively. Consequently, the total circuit count scales as 
	\begin{align}
		N_{\rm total} = \mathcal{O}(N_p^2 N_d^2 + N_c N_h N_p N_d + N_c^2 N_h).
	\end{align}

	\subsubsection{Ancilla-free Hadamard test}
	The conventional Hadamard test employs an ancilla qubit to control the unitaries associated with the Hamiltonian or the ansatz operators, which imposes a large overhead on the quantum circuit depth and complexity. In the context of quantum chemistry, however, the evolution of the system strictly obeys particle-number conservation (i.e., the total number of electrons $N_e$ remains constant). If the unitary operators $U_1$, $\sigma$ and $U_2$ are particle-number preserving, the expressions in $M$ and $V$ can be measured without the ancilla qubit~\cite{Lu2021algorithms,OBrien2021error,Xu2023quantum}.

	First, we demonstrate that $U_1$ and $U_2$ are inherently particle-number-preserving, subject to a minor adjustment. This conclusion is based on several observations: 
	(i) our ansatz is, by construction, particle-number-conserving; 
	(ii) it is evident that $R_{j,k,i}^\dagger$ preserves particle number with respect to the parameters $\varphi, \lambda, a$, and $b$; 
	and (iii) for $\theta$, $R_{j,k,i}$ remains number-conserving upon noting that the $XX+YY$ term in Eq.~(\ref{eq:partial_theta}) can be rewritten using the identity $X \otimes X + Y \otimes Y = 2\text{SWAP} - I \otimes I - Z \otimes Z$. Second, we show that the Hamiltonian can be decomposed as a linear combination of particle-number-preserving unitaries (i.e. $\sigma$). Individual Pauli strings $P_{\nu}$ derived from fermion-to-qubit mappings---such as the Jordan-Wigner, parity, and Bravyi-Kitaev transformations---generally do not preserve particle number. Nevertheless, the Hamiltonian can, in principle, be decomposed into a set of particle-number-preserving unitary operators. 
	For instance, Ref.~\cite{Bespalova2021Hamiltonian} demonstrates that the Hamiltonian can be effectively represented via Hamiltonian operator approximation procedure.

\begin{figure}[htbp]
		\resizebox{\columnwidth}{!}{
			\begin{quantikz}
				\lstick{$\ket{0}$}                 & \gate[2]{U_{\rm GHZ}} & \gate[3]{U_a} & \gate[3]{\sigma} & \gate[3]{U_b} & \gate{R_{\rm Z}(\gamma)} & \gate[2]{U_{\rm GHZ}^\dag} & \meter{} \\
				\lstick{$\ket{0}^{\otimes (N_e-1)}$} &                       &               &                  &               &                          &                            & \bra{0}^{\otimes (N_e-1)} \\
				\lstick{$\ket{0}^{\otimes (N_q-N_e)}$} &                       &               &                  &               &                          &                            & \bra{0}^{\otimes (N_q-N_e)}
			\end{quantikz}
		}
		\caption{Ancilla-free Hadamard test quantum circuit for the estimation of expressions in $M$ and $V$. $U_{\rm GHZ}=\prod_{i=0}^{N_e-2} {\rm CNOT}_{i,i+1} H_0$, which prepares the Greenberger-Horne-Zeilinger (GHZ) state $\ket{\rm GHZ} = \frac{1}{\sqrt{2}} (\ket{0}^{\otimes N_e} + \ket{1}^{\otimes N_e})$.}
		\label{fig:ancilla_free}
	\end{figure}
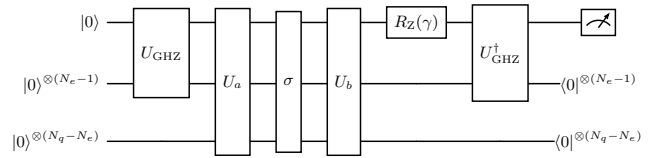

	As illustrated in Fig.~\ref{fig:ancilla_free}, the initial state $\ket{\Psi_0}$ is prepared as the Hartree-Fock state $\ket{1}^{\otimes N_e} \otimes \ket{0}^{\otimes (N_q-N_e)}$, or any other computational basis state with exactly $N_e$ ones. At the end of the circuit, all the qubits are measured in the computational basis. The measurement outcome of the first qubit is recorded as $\mu = \pm 1$ if all other qubits are in the $\ket{0}$ state; otherwise, the shot is discarded. Based on the retained data, the expressions are estimated as $\mathbb{E}[\mu]$. Importantly, this post-selection strategy serves as a powerful error mitigation technique by effectively filtering out unphysical hardware noise and gate errors that cause the system to leak out of the correct symmetry subspace.
	
	\subsubsection{Classical shadow}
	Using the Hadamard test, the number of quantum circuits required to evaluate $M$ and $V$ scales quadratically with the number of parameters $N_p$, the number of nonorthogonal states $N_c$, and the number of decomposed terms $N_d$, while scaling linearly with the number of Hamiltonian terms $N_h$.
	
	By leveraging the classical shadow formalism~\cite{Huang2020Predicting}, we can characterize individual nonorthogonal states and their associated tangent states, thereby circumventing the joint measurements required by the Hadamard test. This strategy reduces the total circuit count from a quadratic to a linear scaling with respect to the number of parameters~\cite{Ren2026Error}. 
	Also, a key advantage of this approach is its independence from the number of Hamiltonian terms $N_h$, which is crucial for quantum chemistry applications where $N_h$ typically scales as $O(N_q^4)$. 
	The total circuit requirement is reduced to 
	\begin{align}
		N_{\rm total} = O(N_p N_d + N_c), 
	\end{align}
	utilizing only $N_q$ qubits without the need for ancilla systems. Furthermore, the circuit remains remarkably shallow, with a depth comparable to that of a single ansatz [approximately $(N_q+1)N_l$] plus the overhead for random unitary transformations. 
	To achieve an estimation precision $\epsilon$ with a confidence level $1-\delta$, the sampling complexity scales as $O(\log(K/\delta)/\epsilon^2)$, where $K$ denotes the number of predicted functions. For observables contain the Hamiltonian, $K$ represents the number of Hamiltonian terms, highlighting the sampling efficiency of the classical shadow framework.
	
\begin{figure}[htbp]
		\centering
		\resizebox{\columnwidth}{!}{
		\begin{quantikz}
			\lstick{$\ket{0}_1$}                   & \gate{H} & \gate[style={dashed, fill=white}]{S} & \ctrl{1} & \qw      & \gate[6]{U_c} & \gate[6]{U_{\rm rand} \in \mathcal{U}} & \meter{} \\
			\lstick{$\ket{0}_2$}                   & \qw      & \qw                                  & \targ{}  & \qw      &                                   &                                                            & \meter{} \\
			\wave{}                                &          &                                      &          &          &                                   &                                                            &          \\
			\lstick{$\ket{0}_{N_e-1}$}             & \qw      & \qw                                  & \qw      & \ctrl{1} &                                   &                                                            & \meter{} \\
			\lstick{$\ket{0}_{N_e}$}               & \qw      & \qw                                  & \qw      & \targ{}  &                                   &                                                            & \meter{} \\
			\lstick{$\ket{0}^{\otimes (N_q-N_e)}$} & \qw      & \qw                                  & \qw      & \qw      &                                   &                                                            & \meter{} 
		\end{quantikz}
	    }
		\caption{The quantum circuit used to extract the interferometric classical shadow of nonorthogonal states ($U_c = U_j$) and tangent states ($U_c = R_{j,k,i}$). Combined with the classical shadow of nonorthogonal states, we can predict $M$ and $V$. The phase gate $S=\ketbra{0}{0} + i\ketbra{1}{1}$. Nearest-neighbor CNOT gates are implemented across the first $N_e$ qubits. The unitary $U_{\rm rand}$ is selected randomly from a fixed ensemble $\mathcal{U}$, which is a Clifford group.}
		\label{fig:classical_shadow}
	\end{figure}
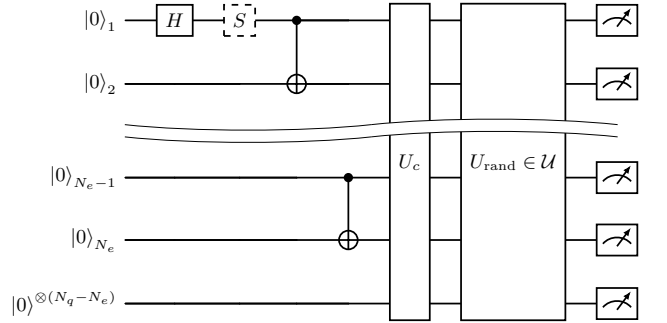

	Following Ref.~\cite{Ren2026Error}, we can estimate $M$ and $V$ without quadratic dependence on $N_p$, $N_c$ and $N_d$. Define quantum states $\ket{\partial_k^i \phi_{j}} = R_{j,k,i} \ket{\Psi_0}$. All the quantities inside $M$ and $V$ can be expressed as the form $\braket{\varphi_i}{\varphi_j}$ or $\bra{\varphi_i} P_\nu \ket{\varphi_j}$, where $\ket{\varphi_i}$ can be $\ket{\phi_j}$ or $\ket{\partial_k^i \phi_{j}}$. To predict these quantities, we first define two interferometric states for each $\ket{\varphi_i}$:
	\begin{align}
		\ket{\varphi_i^R} &= \frac{\ket{0}^{\otimes N_q} +  \ket{\varphi_i}}{\sqrt{2}}, \\
		\ket{\varphi_i^I} &= \frac{\ket{0}^{\otimes N_q} + i\ket{\varphi_i}}{\sqrt{2}}.
	\end{align}
	They can be prepared using the quantum circuit shown in Fig.~\ref{fig:classical_shadow}. To prepare $\ket{\varphi_i^I}$, an additional phase gate is applied. At the end of the circuit, randomized measurement is performed to $\{ \ket{\varphi_i}, \ket{\varphi_i^R}, \ket{\varphi_i^I} \}$ to obtain snapshots $U_{\rm rand}^\dag \ketbra{b}{b} U_{\rm rand}$, where $b$ is the bit string measured in computational basis. Depending on the unitary ensemble $\mathcal{U}$, the classical shadow is calculated as $\mathcal{M}^{-1}(U_{\rm rand}^\dag \ketbra{b}{b} U_{\rm rand})$. If $\mathcal{U} = {\rm Cl}(2^{N_q})$ is the global Clifford group, then $\mathcal{M}^{-1}(X) = (2^{N_q} + 1)X - I$. If $\mathcal{U} = {\rm Cl}(2)^{\otimes N_q}$ is tensor products of the single-qubit Clifford group, i.e. measure each qubit independently in a random Pauli basis, then $\mathcal{M}^{-1}(X) = \otimes_{j=1}^{N_q}(3 X_j - I)$, where $X_j$ are the snapshots of the $j$-th qubit. The density matrices $\ketbra{\varphi_i}{\varphi_i}$, $\ketbra{\varphi_i^R}{\varphi_i^R}$, and $\ketbra{\varphi_i^I}{\varphi_i^I}$ are estimated as the average of all the classical shadows.
	
	With the classical shadow data for $\{ \ket{\varphi_a}, \ket{\varphi_a^R}, \ket{\varphi_a^I} \}_{a=i,j}$, we can now estimate $\{ \braket{\varphi_i}{\varphi_j}, \bra{\varphi_i} P_\nu \ket{\varphi_j} \mid \nu=1,\cdots,N_h \}$. For $\braket{\varphi_i}{\varphi_j}$, we have~\cite{Ren2026Error}
	\begin{align}
		\abs{\braket{\varphi_i}{\varphi_j}}^2 &= \Tr (\ketbra{\varphi_i}{\varphi_i} \ketbra{\varphi_j}{\varphi_j}), \\
		\Re\braket{\varphi_i}{\varphi_j} &= 2 \Tr (\ketbra{\varphi_i^I}{\varphi_i^I} \ketbra{\varphi_j^I}{\varphi_j^I}) - \frac{1}{2} (1 + \abs{\braket{\varphi_i}{\varphi_j}}^2), \\
		\Im\braket{\varphi_i}{\varphi_j} &= 2 \Tr (\ketbra{\varphi_i^I}{\varphi_i^I} \ketbra{\varphi_j^R}{\varphi_j^R}) - \frac{1}{2} (1 + \abs{\braket{\varphi_i}{\varphi_j}}^2).
	\end{align}
	For $\bra{\varphi_i} P_\nu \ket{\varphi_j}$, we have
	\begin{align}
		\bra{\varphi_i} P_\nu \ket{\varphi_j} = \frac{\Tr ( \ketbra{\varphi_j}{\varphi_j} \ketbra{\varphi_i}{\varphi_i} P_\nu )}{\braket{\varphi_j}{\varphi_i}}.  
	\end{align}
	Note that when $i=j$, we simply have $\braket{\varphi_i}{\varphi_i}=1$ and $\bra{\varphi_i} P_\nu \ket{\varphi_i} = \Tr (\ketbra{\varphi_i}{\varphi_i} P_\nu)$.

	\section{Error analysis and resource estimation}
	In this section, we present a systematic error analysis for our algorithm. While the error formalisms for standard VQS and general evolution processes (including ITE) have been established in Refs.~\cite{Li2017efficient} and \cite{Endo2020variational}, we extend this theoretical framework to the NOVQS framework. In our approach, the total error is categorized into two components: algorithmic errors and implementation errors. The former originates from the limited expressive power of the variational ansatz and the discretization of finite time steps (either real or imaginary). The latter is attributed to the statistical shot noise of the quantum computer. In this analysis, the quantum hardware is assumed to be ideal, neglecting hardware imperfections such as gate infidelities and readout errors.
	
	To quantify the algorithmic error, we employ the trace distance, defined as
	\begin{align}
		D(\rho, \sigma) = \frac{1}{2} \text{Tr} \sqrt{(\rho - \sigma)^2}.
	\end{align}
	For pure states $\ket{\psi}$ and $\ket{\phi}$, it can be proven that $D(\ket{\psi}, \ket{\phi})=\sqrt{1-F(\ket{\psi}, \ket{\phi})}$, where the fidelity $F(\ket{\psi}, \ket{\phi}) = \abs{\braket{\psi}{\phi}}^2$.
	In the context of the NOVQS algorithm, the error is evaluated between the normalized variational quantum state $\ket{\tilde{\Phi}(\theta(t))}$ and the normalized target state $\ket{\tilde{\Psi}(t)}$. Considering a total evolution time $T$ with a discrete time step $\delta t$ (either real or imaginary), the simulation is performed over $N_t = T/\delta t$ steps. Consequently, the total algorithmic error at the final step $N_t$ can be expressed as:
	\begin{align}
		\varepsilon = D(\ket{\tilde{\Psi}_{N_t}}, \ket{\tilde{\Phi}_{N_t}})
	\end{align}
	where $\ket{\tilde{\Psi}_{N_t}}$ and $\ket{\tilde{\Phi}_{N_t}}$ represent the normalized target and variational states at the ${N_t}$-th step, respectively. By applying the triangle inequality, we have
	\begin{align}
		&D(\ket{\tilde{\Psi}_{N_t}}, \ket{\tilde{\Phi}_{N_t}})  \notag\\
		&\leq  D(\ket{\tilde{\Psi}_{N_t}}, \mathcal{E}_{N_t}\ket{\tilde{\Phi}_{N_t-1}}) + D(\mathcal{E}_{N_t}\ket{\tilde{\Phi}_{N_t-1}}, \ket{\tilde{\Phi}_{N_t}}) \notag\\
		&= D(\mathcal{E}_{N_t}\ket{\tilde{\Psi}_{N_t-1}}, \mathcal{E}_{N_t}\ket{\tilde{\Phi}_{N_t-1}}) + D(\mathcal{E}_{N_t}\ket{\tilde{\Phi}_{N_t-1}}, \ket{\tilde{\Phi}_{N_t}}) \notag\\
		&\leq D(\ket{\tilde{\Psi}_{N_t-1}}, \ket{\tilde{\Phi}_{N_t-1}}) + D(\mathcal{E}_{N_t}\ket{\tilde{\Phi}_{N_t-1}}, \ket{\tilde{\Phi}_{N_t}}) \notag\\
		&\leq D(\ket{\tilde{\Psi}_{N_t-2}}, \ket{\tilde{\Phi}_{N_t-2}}) + \sum_{i=N_t-1}^{N_t} D(\mathcal{E}_i\ket{\tilde{\Phi}_{i-1}}, \ket{\tilde{\Phi}_i}) \notag\\
		&\cdots \notag\\
		&\leq  D(\ket{\tilde{\Psi}_1}, \ket{\tilde{\Phi}_1}) + \sum_{i=2}^{N_t} D(\mathcal{E}_i\ket{\tilde{\Phi}_{i-1}}, \ket{\tilde{\Phi}_i}) \notag\\
		&\leq  D(\ket{\tilde{\Psi}_0}, \ket{\tilde{\Phi}_0}) + \sum_{i=1}^{N_t} D(\mathcal{E}_i\ket{\tilde{\Phi}_{i-1}}, \ket{\tilde{\Phi}_i})
	\end{align}
	where $\mathcal{E}_i$ denotes the exact evolution map at $i$-th time step. For real-time evolution, $\mathcal{E}$ is unitary and thus acts as an isometry that preserves the trace distance. In contrast, imaginary-time evolution is contractive, effectively reducing the distance between any two states. Consequently, the inequality in the fourth line holds universally for both types of quantum dynamics. In the final line, $\ket{\tilde{\Psi}_0}$ represents the exact initial state. Assuming perfect initial state preparation on the quantum computer, we have $\ket{\tilde{\Psi}_0} = \ket{\tilde{\Phi}_0} = \ket{\Psi_0}$, such that the initial error $D(\ket{\tilde{\Psi}_0}, \ket{\tilde{\Phi}_0}) = 0$. Consequently, we obtain
	\begin{align}
		\varepsilon &\leq \sum_{i=1}^{N_t} D(\mathcal{E}_i\ket{\tilde{\Phi}_{i-1}}, \ket{\tilde{\Phi}_i}) \notag\\
		&\leq \sum_{i=1}^{N_t} D(\mathcal{E}_i\ket{\tilde{\Phi}_{i-1}}, \ket{\tilde{\Phi}_i^{(0)}}) + \sum_{i=1}^{N_t} D(\ket{\tilde{\Phi}_i^{(0)}}, \ket{\tilde{\Phi}_i})    
	\end{align}
	where $\ket{\tilde{\Phi}_i^{(0)}}$ denotes the exact state optimized from exact $M$ and $V$ at the $i$-th time step. The first term, $\varepsilon_A = \sum_i \varepsilon_{A,i}$, accounts for imperfect ansatz and finite time step. The second term, $\varepsilon_I = \sum_i \varepsilon_{I,i}$, is introduced by the noise when running the quantum computer.

	\subsection{Algorithmic error}
	To quantify the algorithmic error, we first expand both the exact and variational wavefunctions in a Taylor series with respect to time $t$:
	\begin{align}
		\mathcal{E}_i\ket{\tilde{\Phi}_{i-1}} &= \ket{\tilde{\Phi}_{i-1}(t)} + \delta t \frac{d}{dt}\ket{\tilde{\Phi}_{i-1}(t)} \notag\\
		&+ \frac{1}{2} \delta t^2 \frac{d^2}{dt^2} \ket{\tilde{\Phi}_{i-1}(t)} + \mathcal{O} (\delta t^3), \notag\\
		\ket{\tilde{\Phi}_i^{(0)}} &= \ket{\tilde{\Phi}_{i-1}(\theta(t))} + \delta t \sum_j \dot{\theta}_j \frac{\partial}{\partial \theta_j} \ket{\tilde{\Phi}_{i-1}(\theta(t))} \notag\\
		&+ \frac{1}{2} \delta t^2  \sum_{j,k} \dot{\theta}_j \dot{\theta}_k \frac{\partial^2}{\partial \theta_j \partial \theta_k} \ket{\tilde{\Phi}_{i-1}(\theta(t))} + \mathcal{O}(\delta t^3).
		\label{eq:tylor}
	\end{align}
	The trace distance between the exact and variational states can be rewritten as $\varepsilon_{A,i} = \sqrt{\braket{\delta \Phi_i}{\delta \Phi_i} - \abs{\braket{\tilde{\Phi}_i^{(0)}}{\delta \Phi_i}}^2}$, where the error vector $\ket{\delta \Phi_i} = \mathcal{E}_i\ket{\tilde{\Phi}_{i-1}} - \ket{\tilde{\Phi}_i^{(0)}}$. The first- and second-order error vectors are 
	\begin{align}
		\ket{\delta \Phi_i^{(1)}} &= \frac{d}{dt}\ket{\tilde{\Phi}_{i-1}(t)} - \sum_j \dot{\theta}_j \frac{\partial}{\partial \theta_j} \ket{\tilde{\Phi}_{i-1}(\theta(t))}, \notag\\
		\ket{\delta \Phi_i^{(2)}} &= \frac{1}{2} \left( \frac{d^2}{dt^2} \ket{\tilde{\Phi}_{i-1}(t)} - \sum_{j,k} \dot{\theta}_j \dot{\theta}_k \frac{\partial^2}{\partial \theta_j \partial \theta_k} \ket{\tilde{\Phi}_{i-1}(\theta(t))} \right),     
	\end{align}
	such that $\ket{\delta \Phi_i} = \delta t \ket{\delta \Phi_i^{(1)}} + \delta t^2 \ket{\delta \Phi_i^{(2)}} + \mathcal{O}(\delta t^3)$.
	Therefore, the algorithmic error at each step is upper bounded as 
	\begin{align}
		\varepsilon_{A,i} = \sqrt{\Delta_2 \delta t^2 + \Delta_3 \delta t^3 + \mathcal{O}(\delta t^4)} \lesssim \sqrt{\Delta_2} \delta t + \sqrt{\Delta_3 \delta t} \delta t,
	\end{align}
	where $\Delta_2 = \braket{\delta \Phi_i^{(1)}}{\delta \Phi_i^{(1)}} - \abs{\braket{\tilde{\Phi}_{i-1}(\theta(t))}{\delta \Phi_i^{(1)}}}^2$ characterizes the ansatz imperfect error and $\Delta_3 = 2\Re \braket{\delta \Phi_i^{(1)}}{\delta \Phi_i^{(2)}}$ represents the finite-time discretization error. The total algorithmic error for the entire evolution time $T$ is
	\begin{align}
		\varepsilon_A \lesssim \sqrt{\Delta_2^{\rm max}} T + \sqrt{\Delta_3^{\rm max} \delta t} T,
	\end{align}
	where $\Delta_2^{\rm max}$ and $\Delta_3^{\rm max}$ denote the maximum values of $\Delta_2$ and $\Delta_3$ across the $N_t$ time steps. 
	Assuming the variational ansatz can express the time evolution exactly, i.e. $\ket{\delta \Phi_i^{(1)}}=0$ for all $i$, we have $\Delta_2^{\rm max} = 0$. Then we get $\varepsilon_A \lesssim \sqrt{\Delta_3^{\rm max} \delta t} T$. 
    To ensure the algorithmic error within $\varepsilon_A$, the time step should be set as
	\begin{align}
		\delta t \approx \frac{\varepsilon_A^2}{\Delta_3^{\rm max} T^2},
	\end{align}
	and the corresponding total time steps
	\begin{align}
		N_t \approx \Delta_3^{\rm max} T^3 / \varepsilon_A^2.
	\end{align}
	
	\subsection{Implementation error}
	The implementation error $\varepsilon_{I,i}$ originates from the imprecise estimation of $M$ and $V$ at the $i$-th step. 
	Let $M'$ and $V'$ denote the estimated quantum Fisher information matrix and force vector, respectively, from which the 
	approximate derivative is computed as $\dot{\theta}' = (M')^{-1} V'$. The associated discrepancies from their exact counterparts are defined as $\delta M = M' - M$, $\delta V = V' - V$ and $\delta\dot{\theta} = \dot{\theta}' - \dot{\theta}$. Analogous to Eq.~(\ref{eq:tylor}), the variational state at the $i$-th step can be expanded as
	\begin{align}
		\ket{\tilde{\Phi}_i} = \ket{\tilde{\Phi}_{i-1}(\theta(t))} + \delta t \sum_j \dot{\theta}'_j \frac{\partial}{\partial \theta_j} \ket{\tilde{\Phi}_{i-1}(\theta(t))} + \mathcal{O}(\delta t^2).
	\end{align}
	Defining the first-order error vector 
	\begin{align}
		\ket{\delta \Phi'_i} = \delta t \sum_j \delta\dot{\theta}_j \frac{\partial}{\partial \theta_j} \ket{\tilde{\Phi}_{i-1}(\theta(t))},
	\end{align}
	the implementation error can be quantified as
	\begin{align}
		\varepsilon_{I,i} &= \sqrt{\braket{\delta\Phi'_i}{\delta\Phi'_i} - \abs{\braket{\tilde{\Phi}_i^{(0)}}{\delta \Phi'_i}}^2} \notag\\
		&= \sqrt{\delta \dot{\theta}^T M \delta \dot{\theta} \delta t^2 + \mathcal{O}(\delta t^3)},
	\end{align}
	where $M$ is defined in Eq.~(\ref{eq:rte-M}). 
	This local error can be upper bounded by 
	\begin{align}
		\varepsilon_{I,i} \lesssim \sqrt{\norm{M}} \norm{\delta\dot{\theta}} \delta t.
	\end{align}

	From the defination of $\delta\dot{\theta}$, we have
	\begin{align}
		\delta\dot{\theta} &= (M')^{(-1)} V' - M^{-1} V \notag\\
		&=[M(I + M^{-1} \delta M)]^{-1} (V+\delta V) - M^{-1}V \notag\\
		&=(I + M^{-1} \delta M)^{-1} M^{-1} (V+\delta V) - M^{-1}V \notag\\
		&\approx (I - M^{-1} \delta M) (M^{-1}V + M^{-1}\delta V) - M^{-1}V \notag\\
		&\approx M^{-1}\delta V - M^{-1} \delta M M^{-1}V.
	\end{align}
	Here, we assume the condition number $\kappa(M)$ is not large and $\norm{\delta M} \ll \norm{M}$, which gives $\norm{M^{-1} \delta M} \leq \norm{M^{-1}} \norm{\delta M} =\kappa(M) \norm{\delta M} / \norm{M} \ll 1$.
	Applying the triangle inequality yields the norm bound
	\begin{align}
		\norm{\delta\dot{\theta}} \leq \norm{M^{-1}} \norm{\delta V} + \norm{M^{-1}}^2 \norm{V} \norm{\delta M}.
	\end{align}
	
	Let indices $k$ and $q$ correspond to the $j$- and $j'$-th ansatz components, respectively. We assume an ideal quantum processor with no machine errors, considering only the contribution of shot noise. With $N_s$ measurements allocated to each term, 
	\begin{align}
		\norm{\delta M} \leq \frac{\alpha \sqrt{\sum_{k,q} m_{k,q}^2}}{\sqrt{N_s}},
	\end{align}
	where $\alpha=1/\braket{\Psi}{\Psi}$ acts as the normalization factor and $m_{k,q}^2 = \sum_{i,i'} \abs{f_{j,k,i}^* f_{j',q,i'}}^2 + 2\alpha^2 (\sum_{l,i}\abs{c_l f^*_{j,k,i}})^2 (\sum_{l,i}\abs{c^*_l f_{j',q,i}})^2$. 
	By an analogous derivation, the perturbation on the force vector is bounded by
	\begin{align}
		\norm{\delta V} \leq \frac{\alpha \sqrt{\sum_{k,q} v_{k,q}^2}}{\sqrt{N_s}},
	\end{align}
	where $v_k^2 = \sum_{i,\nu,l} \abs{f^*_{j,k,i} h_{\nu} c_l}^2 + 2\alpha^2 (\sum_{l,\nu,l'} \abs{c_l^* h_{\nu} c_{l'}})^2 (\sum_{i,l} \abs{f^*_{j,k,i} c_l})^2$.
	Substituting these results back into the $\norm{\delta \dot{\theta}}$ bound gives
	\begin{align}
		\norm{\delta \dot{\theta}} \leq \frac{\alpha \norm{M^{-1}} \Delta}{\sqrt{N_s}},
	\end{align}
	where $\Delta =  \sqrt{\sum_{k,q} v_{k,q}^2} + \norm{M^{-1}} \norm{V} \sqrt{\sum_{k,q} m_{k,q}^2}$. 
	The local implementation error is thus
	\begin{align}
		\varepsilon_{I,i} \lesssim \sqrt{\norm{M}} \frac{\alpha \norm{M^{-1}} \Delta}{\sqrt{N_s}} \delta t.
	\end{align}
	Finally, the total implementation error can be represented as
	\begin{align}
		\varepsilon_I \lesssim \sqrt{\norm{M}_{\rm max}} \frac{\alpha_{\rm max} \norm{M^{-1}}_{\rm max} \Delta_{\rm max}}{\sqrt{N_s}} T,
	\end{align}
	where where the subscript 'max' denotes their maximum value across $N_t$ steps. This result reveals that the implementation error is sensitive to the condition number of $M$ and the normalization factor of the variational wavefunction $\alpha$. To achieve a permissible error of $\epsilon_I$, the required measurement number is
	\begin{align}
		N_s \approx \alpha_{\rm max}^2 \norm{M}_{\rm max} \norm{M^{-1}}_{\rm max}^2 \Delta_{\rm max}^2 T^2 / \varepsilon_I^2.
	\end{align}

	\subsection{Resource estimation}
	Through the analyses of the algorithmic and implementation errors, we obtain conservative estimates of the total number of time steps and measurement shots. At each time step, when classical shadows are employed, the number of quantum circuits to be measured is $\mathcal{O}(N_p N_d+N_c)$. Therefore, for target algorithmic and implementation errors of $\varepsilon_A$ and $\varepsilon_I$, respectively, the total number of measurements is estimated as
	\begin{align}
		&N_t N_s N_c\notag\\
		\sim&  \frac{\alpha_{\rm max}^2 \norm{M}_{\rm max} \norm{M^{-1}}_{\rm max}^2 \Delta_{\rm max}^2 \Delta_3^{\rm max} T^5}{\varepsilon_A^2 \varepsilon_I^2} (N_p N_d + N_c).
	\end{align}
	In conclusion, we analyze the error of our algorithm and determine the quantum resource required to achieve a target precision. We demonstrate that the algorithm scales polynomially with respect to the evolution time $T$, algorithmic error $\varepsilon_A$, implementation error $\varepsilon_I$, and the number of parameters $N_p$.

	\section{Results}
	
	\subsection{Performance comparison}\label{sec:comparison}
	In this section, we benchmark the performance of VQS and NOVQS in simulating both RTE and ITE. We demonstrate that NOVQS, even leveraging an ensemble of single-layer PQCs, can match or even outperform VQS implemented with significantly deeper circuits.

	\subsubsection{Real time evolution}
	
	First, we simulate the real time evolution using both VQS and NOVQS. To evaluate the algorithmic performance, we compute the fidelity between the variational wavefunction and the exact time-evolved state. As a benchmark, we consider an $\mathrm{H}_4$ chain molecule at a relatively large interatomic distance of $R = 2.0 \, \text{\AA}$. To explore the expressivity limits of both methods, we set $\ket{\Psi_0}$ to be the uniform superposition state, $\ket{+}^{\otimes N_q}$, and randomly sample the parameters from a uniform distribution over the interval $[-1, 1]$. For NOVQS, the initial coefficients $c_j(t=0)$ are $1$. To update the variational parameters, we employ the fourth-order Runge-Kutta (RK4) method, which significantly suppresses the accumulated integration error compared to the basic Euler method illustrated in Fig.~\ref{fig:schematic}. Additionally, to avert numerical instability arising from machine precision, we regularize the $M$ matrix via a diagonal shift $M \to M + \epsilon_{\rm reg} I$ (setting $\epsilon_{\rm reg} = 10^{-8}$ in our simulations).

	In this stretched bond-length regime, the $\mathrm{H}_4$ molecule exhibits strong electron correlation, making real time simulations challenging. Nevertheless, both VQS and NOVQS can capture the time evolved state with exceptionally high precision. This demonstrates the powerful expressivity of our proposed ansatz in describing fermionic systems such as quantum chemical molecules.
	As shown in the figure, increasing the circuit depth of VQS from $N_l=1$ to $8$ layers steadily improves the fidelity, culminating in an accuracy on the order of $10^{-4}$. In contrast, the NOVQS protocol achieves an exceptional accuracy of $10^{-8}$ by utilizing $N_c=8$ quantum circuits of merely a single layer. Notably, this precision effectively saturates the limit imposed by the regularization factor ($\epsilon_{\rm reg} = 10^{-8}$). Furthermore, we performed analogous simulations for the molecule at a near-equilibrium bond length of $R = 0.74 \, \text{\AA}$, yielding highly consistent results.

	\begin{figure}[htbp]
		\centering
		\includegraphics[width=\linewidth]{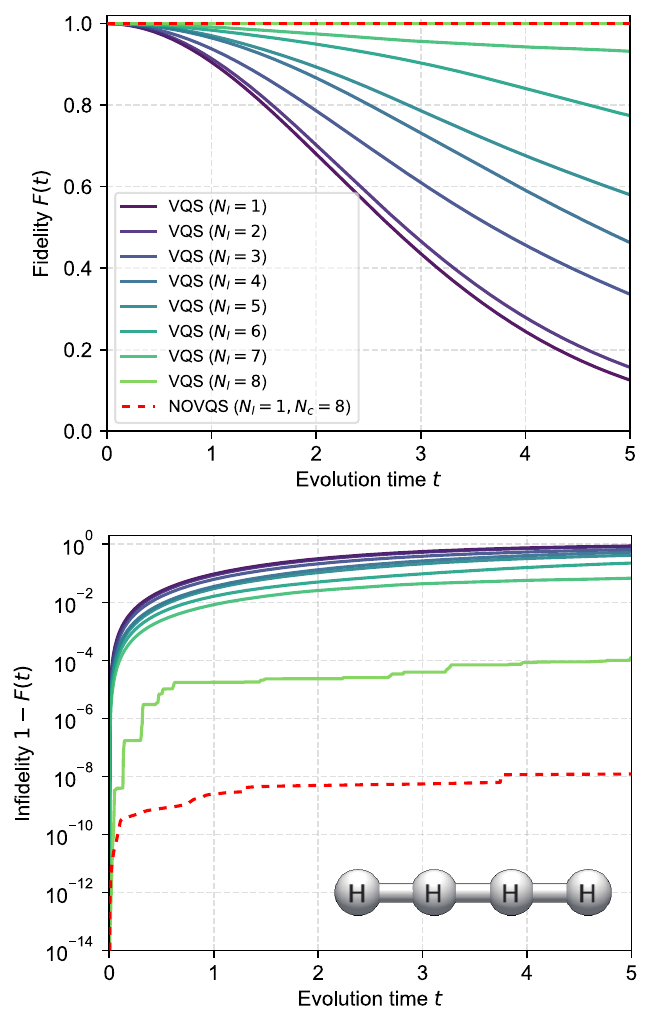}
		\caption{Variational real time evolution using VQS and NOVQS. The simulated model is an $\mathrm{H}_4$ chain molecule with an interatomic distance of $2.0\, \text{\AA}$ in the STO-3G basis set. The initial state is prepared as the uniform superposition state $\ket{+}^{\otimes N_q}$, with the parameters drawn uniformly from the range $[-1, 1]$. For NOVQS, all coefficients are initialized to $1$. (Top panel) The fidelity of the variational wavefunction with respect to the exact solution, $F(t) = \abs{\braket{\Psi(t)}{\Phi(\theta(t))}}^2$. The time step is set to $\delta t = 0.005$. (Bottom panel) The corresponding infidelity, $1-F(t)$, shown on a logarithmic scale. The red dashed line represents the fidelity of the NOVQS using $N_c=8$ single-layer parameterized quantum circuits, while the solid lines depict the VQS fidelity for circuit layers ranging from $1$ to $8$.}
		\label{fig:RTE-H4-comparison}
	\end{figure}

	\subsubsection{Imaginary time evolution}
	
	While the ansatz is designed based on quantum simulation, it has been widely studied that the integral of time evolved operators over time $t$ composes a energy filter, e.g. sinc~\cite{Neuhauser1990bound} and Gaussian~\cite{Wall1995extraction} function, which makes it also well suited for imaginary time evolution. 
	Note that the discrete approximation of such integrals naturally leads to a linear combination of time-evolved operators, which aligns with the structure of our proposed ansatz.

	When simulating imaginary time evolution (ITE), we use the energy as the primary metric to evaluate the quality of the variational wavefunction. Provided that the initial state has a non-zero overlap with the true ground state, the ITE process naturally drives the system toward the lowest-energy ground state. This robust cooling property has established ITE as the foundational computational backend for a wide array of quantum Monte Carlo methods.

\begin{figure}[htbp]
		\centering
		\includegraphics[width=\linewidth]{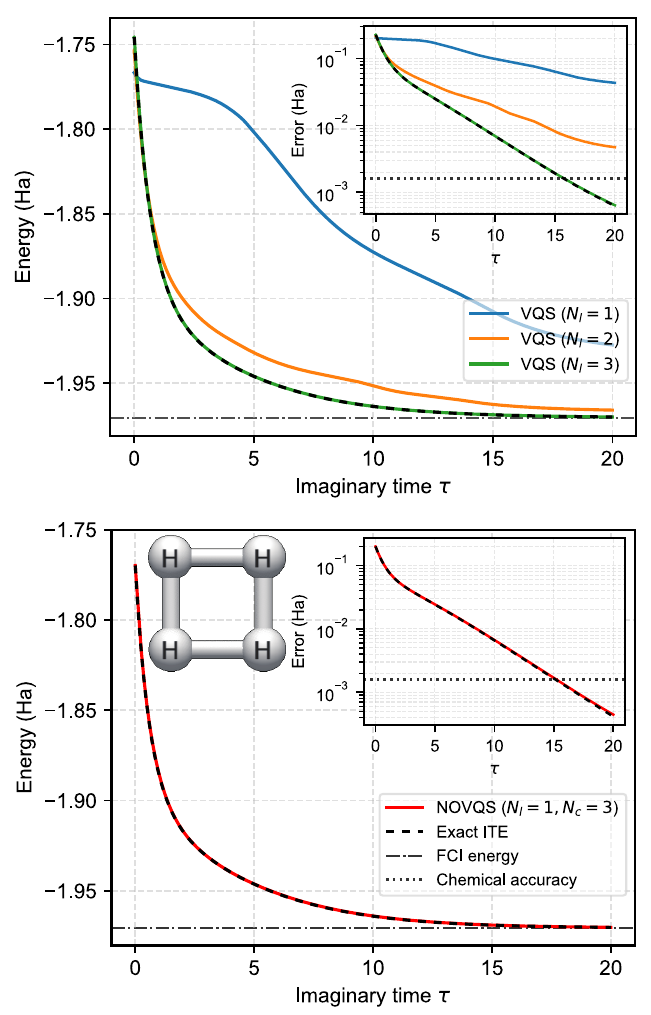}
		\caption{Variational imaginary time evolution using VQS (top panel) and NOVQS (bottom panel).The simulated model is an $\mathrm{H}_4$ molecule in a square geometry with an equilibrium bond length of $1.28\, \text{\AA}$ in the STO-3G basis set. The simulation starts from the Hartree-Fock state with a small random perturbation (at most $\pi/50$) applied to all circuit parameters. For NOVQS, all coefficients are initialized to $1$. The time step is set to $\delta \tau = 0.005$. Insets show the absolute energy error relative to the exact ITE on a logarithmic scale. The dashed black line represents the exact ITE trajectory, while the dash-dotted and dotted lines denote the FCI energy and chemical accuracy, respectively.}
		\label{fig:ITE-H4-comparison}
	\end{figure}

	We consider the square $\mathrm{H}_4$ molecule, which is a demanding benchmark system characterized by strong multi-reference character and competing ground states. In the STO-3G basis set, the equilibrium bond length of the square $\mathrm{H}_4$ molecule is located at $R = 1.28 \, \text{\AA}$. We select the Hartree-Fock state as the initial state, which exhibits a substantial overlap with the true ground state, yielding a fidelity of $0.454$. The variational parameters are initialized with small random perturbations around zero, sampled uniformly from the interval $[-\pi/50,\pi/50]$. The initial coefficients are set to $1$ for NOVQS. In Fig.~\ref{fig:ITE-H4-comparison}, we present the simulation results of ITE using VQS and NOVQS, respectively. Both methods perform exceptionally well in tracking the energy trajectory of the exact ITE, successfully converging to chemical accuracy after $\tau=15$. While the standard VQS requires a circuit depth of $N_l=3$ layers to perform optimally, the NOVQS protocol achieves an equivalent level of accuracy using an ensemble of merely $N_c=3$ independent single-layer circuits. This not only demonstrates the effectiveness of our PQCs for ITE simulations, but also highlights a critical advantage of NOVQS: it can deliver a performance on par with deep VQS architectures by leveraging multiple extremely shallow circuits. In other words, with limited quantum circuit depth, NOVQS can outperform single-circuit-based VQS in both RTE and ITE simulations by employing several shallow sub-circuits in parallel.

	\subsection{$N_c$-$N_l$ trade-off}
	In Sec.~\ref{sec:comparison}, it is shown that $n$ single-layer quantum circuits exhibit performance comparable to that of a single $n$-layer quantum circuit. To further substantiate this observation, we benchmark the performance of quantum chemistry simulations using different numbers of PQCs ($N_c$) and circuit layers ($N_l$). 
	
	In principle, our algorithm allows different circuits to have different depths. For simplicity, however, we use the same number of layers for each PQC in this benchmark. Specifically, we simulate the real-time evolution of the linear $\mathrm{H}_4$ and $\mathrm{H}_5$ molecular chains. For $\mathrm{H}_4$ molecule, we initialize the coefficients as $c_j=1$ and set the gate parameters $\theta$, $\varphi$, and $\lambda$ to zero, such that all configurations start from the uniform superposition state $\ket{+}^{\otimes N_q}$. For $\mathrm{H}_5$ molecule, the coefficients are initialized to $1$, the gate parameters are independently sampled from the interval $[-\pi/50,\pi/50]$, with the Hartree-Fock state used as the reference state. We use the infidelity at the final evolution time $T$, defined as $1-\left|\braket{\Psi(T)}{\Phi(\boldsymbol{\theta}(T))}\right|^2$,
	as the performance metric.
	
\begin{figure}[htbp]
		\centering
		\includegraphics[width=\linewidth]{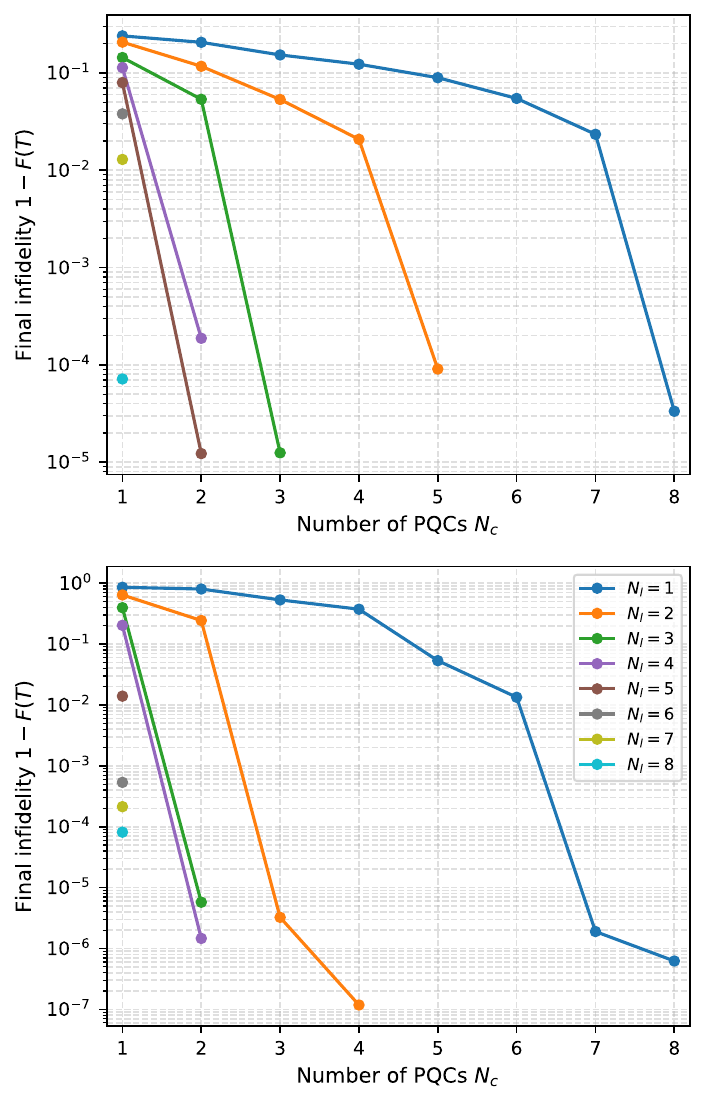}
		\caption{Final infidelities of NOVQS for different numbers of PQCs ($N_c$) and circuit layers ($N_l$). Note that when $N_c=1$, NOVQS reduces to VQS. (Top panel) Results for a linear $\mathrm{H}_4$ molecular chain with an interatomic spacing of $2.0\,\text{\AA}$, described using the STO-3G basis set. The initial state is $\ket{+}^{\otimes N_q}$, with the coefficients initialized to $1$ and the gate parameters initialized to $0$. (Bottom panel) Results for a linear $\mathrm{H}_5$ molecular chain with an interatomic spacing of $2.0\,\text{\AA}$, described using the STO-3G basis set. The initial state is the Hartree-Fock state, with the coefficients initialized to $1$ and the gate parameters randomly perturbed around $0$ with a maximum magnitude of $\pi/50$. In both cases, the total evolution time is $T=5.0$, and the time step is $\delta t=0.005$. The final infidelity is defined as $1-\left|\braket{\Psi(T)}{\Phi(\theta(T))}\right|^2$.}
		\label{fig:Nc-Nl_trade-off}
	\end{figure}

	The results are shown in Fig.~\ref{fig:Nc-Nl_trade-off}, which reveal a clear trade-off between $N_c$ and $N_l$. For example, as shown in the top panel, achieving a final infidelity of approximately $10^{-4}$ requires eight single-layer PQCs, five two-layer PQCs, three three-layer PQCs, or two four-layer PQCs. Similarly, in the bottom panel, achieving a final infidelity of approximately $10^{-6}$ requires seven single-layer PQCs, three two-layer PQCs, two three-layer PQCs, or two four-layer PQCs. We conclude that different configurations achieving comparable accuracy require approximately the same total circuit depth, which scales as $\mathcal{O}(N_c N_l)$, providing useful guidance for practical implementations. Therefore, when circuit depth is constrained by hardware, multiple quantum circuits can be employed to enhance the expressivity of the variational ansatz.
	
	\subsection{Molecular benchmarks}
	In this section, we further demonstrate the strong performance of NOVQS in practical applications, including the real-time simulation of spatial-orbital occupations and the calculation of molecular potential energy curves.
	
\begin{figure}[htbp]
		\centering
		\includegraphics[width=\linewidth]{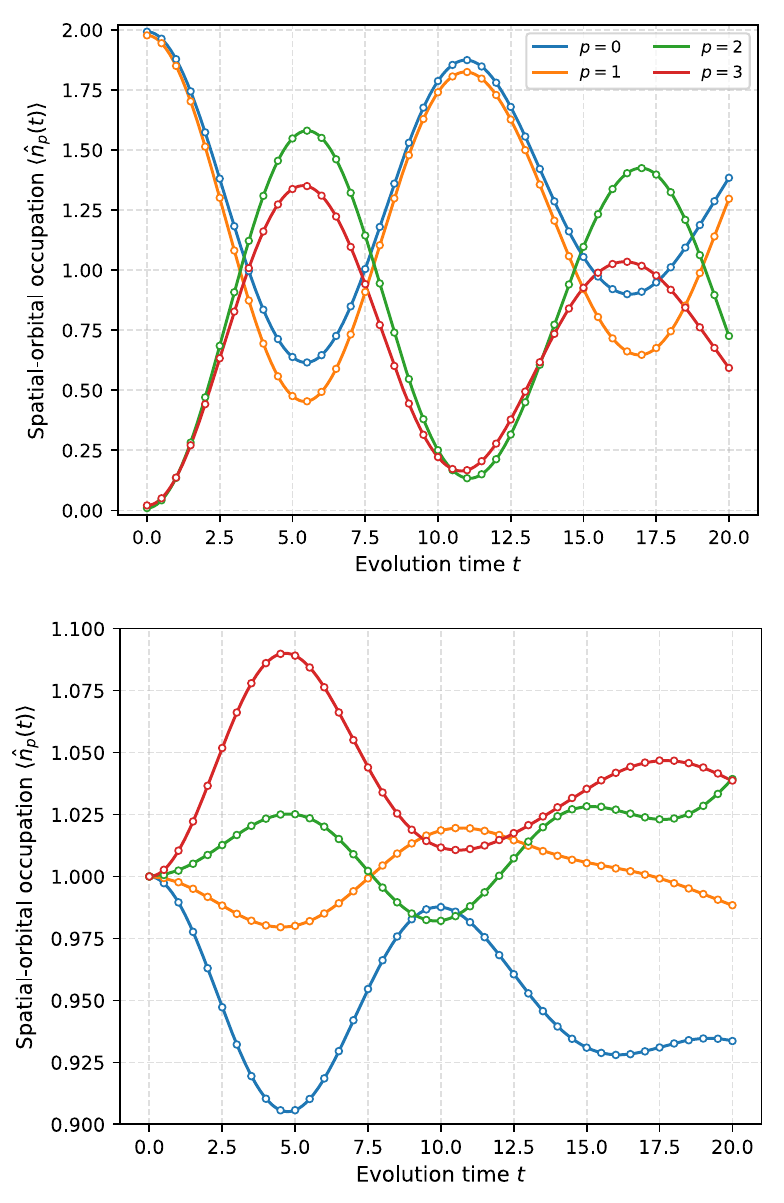}
		\caption{NOVQS results for the time-dependent spatial-orbital occupation $\mean{\hat{n}_p(t)}$. The simulated model is a linear $\mathrm{H}_4$ molecular chain with interatomic distance $2.0\,\text{\AA}$, and $p$ indicates the $p$-th spatial-orbit. (Top panel) The initial state is the Hartree-Fock state with the initial parameters perturbated by at most $\pi/50$. (Bottom Panel) The initial state is the uniform superposition state. For both cases, the number of PQCs is $3$ and the circuit layer is $3$. The total evolution time is $T=20$ and the time step is $\delta t=0.005$. The solid lines represent the NOVQS results, while the open circles indicate the exact reference values.}
		\label{fig:occupations}
	\end{figure}

	First, we use NOVQS to simulate the time-dependent electron occupations of the spatial orbitals in a hydrogen chain up to $T=20$, as shown in Fig.~\ref{fig:occupations}. In Fig.~\ref{fig:occupations}(a), the evolution starts from the Hartree-Fock state. The spatial orbitals labeled by $p=0$ and $1$ are the two lowest-energy orbitals and are therefore initially doubly occupied. In Fig.~\ref{fig:occupations}(b), the system is initialized in the uniform superposition state, for which the average initial occupation of each spatial orbital is one. The results show that three PQCs, each consisting of three layers, reproduce the exact results with excellent accuracy throughout the entire evolution.
	
\begin{figure}[htbp]
		\centering
		\includegraphics[width=\linewidth]{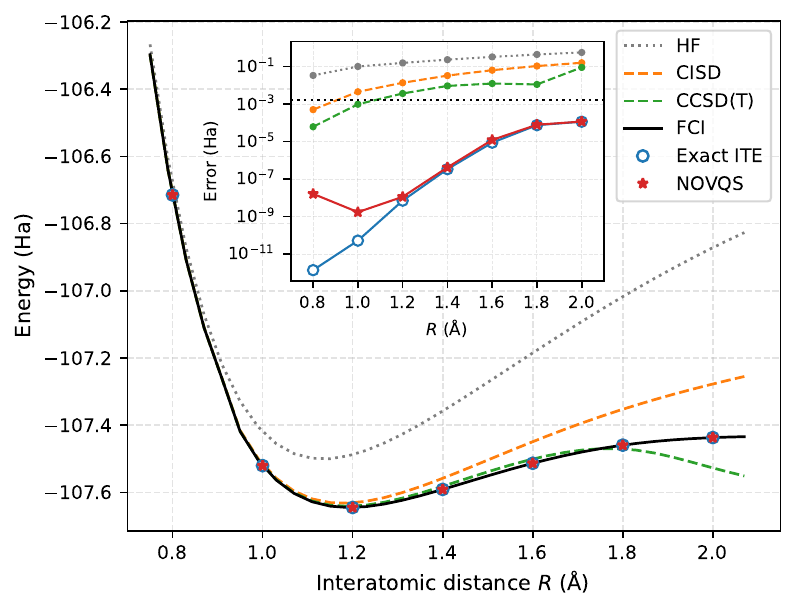}
		\caption{NOVQS calculation of the ground-state energy of the nitrogen molecule. We employ the STO-3G basis set and freeze the four lowest-energy doubly occupied spatial molecular orbitals, resulting in a $\mathrm{N_2\,(6e,6o)}$ model requiring 12 qubits. Starting from the Hartree-Fock state, we randomly perturb the initial gate parameters around zero by at most $\pi/50$ and perform imaginary-time evolution up to $\tau_{\max}=20$. The NOVQS ansatz consists of 18 single-layer PQCs, and its parameters are updated using the Euler method with a time step of $\delta \tau=0.01$. The red stars and open blue circles represent the energies obtained with NOVQS and exact imaginary-time evolution, respectively, at the final imaginary time $\tau_{\max}$. The HF, CISD, CCSD(T), and FCI results are also shown for comparison.}
		\label{fig:energy_curve}
	\end{figure}

	In Fig.~\ref{fig:energy_curve}, we employ NOVQS to determine the energy curve of the $\mathrm{N}_2$ molecule. Starting from an approximate Hartree-Fock state, we evolve the system in imaginary time. Because the initial state has a sufficiently large overlap with the ground state, the evolved state is projected onto the ground state after a sufficiently long imaginary time. For all interatomic distances, we fix the total imaginary time to $\tau_{\max}=20$. Since our objective here is to determine the ground-state energy rather than to accurately track the entire imaginary-time evolution trajectory, we use the Euler method to update the parameters with a relatively large time step of $\delta\tau=0.01$. Nevertheless, the NOVQS trajectories remain in excellent agreement with the exact ITE results. Using 18 single-layer PQCs, NOVQS achieves high accuracy, with the final energy errors remaining below the chemical accuracy threshold.
	
	At short interatomic distances, the energy error saturates at approximately $10^{-8}$. This numerical floor arises from the regularization applied to the matrix $M$ to improve numerical stability, namely, $M\rightarrow M+ 10^{-8} I$. In the dissociation regime, the ground state gradually changes from a predominantly Hartree-Fock, single-reference state to a multireference state. Consequently, the ground-state fidelity of the initial reference state decreases as the interatomic distance increases. For a fixed imaginary-time evolution duration, this reduced initial overlap leads to a deterioration in the final accuracy.
	
	\section{Conclusions and outlook}
	
	In this work, we establish the NOVQS framework for quantum chemistry simulations. We construct the variational wavefunction as a linear combination of nonorthogonal quantum states generated by PQCs. A similar approach has been applied to VQE for estimating ground-state energies, and here we further extend it to VQS.
	Based on McLachlan’s variational principle, we derive the EOMs governing the parameters of an unnormalized wavefunction under both real- and imaginary-time evolution. These equations allow us to jointly update the parameters of the PQCs and their corresponding complex coefficients. Inspired by fermionic swap network~\cite{fSim2018Kivlichan}, we design a hardware-friendly, shallow, and efficient PQC tailored to the simulation of second-quantized fermionic systems. We present several protocols for measuring the matrices and vectors $M$ and $V$ in the EOMs, including the standard and ancilla-free Hadamard tests and a classical-shadow protocol. In particular, the classical-shadow protocol requires no ancilla qubits and substantially reduces both the depth and the number of measurement circuits. We also provide a detailed error analysis and resource estimation for the NOVQS algorithm. 
	As a further application, we show in Appendix~\ref{app:QPE} that NOVQS can serve as a subroutine for single-ancilla quantum phase estimation.
	In Appendix~\ref{app:preparation}, we present a protocol for directly implementing the NOVQS wavefunction within a single quantum circuit, thereby extending its range of applications.

	We performed a variety of numerical simulations of NOVQS, including both real- and imaginary-time dynamics. These results demonstrate the effectiveness of the proposed ansatz and the flexibility of the variational subspace spanned by multiple parameterized quantum states. We compared the performance of VQS and NOVQS for both types of evolution. Although both methods accurately capture the evolution trajectories, VQS requires relatively deep quantum circuits, whereas NOVQS can achieve comparable performance using only a collection of single-layer circuits.
	We further observe a trade-off between circuit depth and the number of circuits in NOVQS. Compared with conventional VQS, NOVQS can systematically enhance its expressive power under circuit-depth constraints by employing multiple PQCs. For the same total number of variational parameters, VQS and NOVQS require the estimation of the same number of matrix and vector elements in the equations of motion. However, NOVQS incurs additional measurement overhead when estimating each element. When classical shadows are used, the total number of measurement circuits increases from $\mathcal{O}(N_pN_d)$ to $\mathcal{O}(N_pN_d+N_c)$. This increase represents an inherent cost of using multiple quantum circuits to jointly represent a single quantum state. It is worth noting that although the total number of circuits increases, the depth of each circuit is substantially reduced.
	As illustrative applications, we used NOVQS to simulate the time-dependent spatial-orbital occupations of hydrogen chains, obtaining excellent agreement with the exact results over extended evolution times. We also employed imaginary-time evolution to determine the ground-state energy of the nitrogen molecule, achieving similarly high accuracy. It should be emphasized that the performance of this approach depends on the availability of a suitable reference state. In the dissociation region, the Hartree–Fock state may no longer provide a good reference. The preparation of high-quality initial reference states remains an active area of research~\cite{Lee2023Evaluating}.

	Promising directions for future research include experimental demonstrations on early fault-tolerant quantum computers, the development of error-correction schemes tailored to the algorithm, and applications of NOVQS to electronic structure calculations and dynamical simulations of larger systems. Numerical evidence suggests that the number of parameters required by TNS for dynamical simulations can grow exponentially with evolution time~\cite{Lin2021Real}. Meanwhile, NQS used to determine the ground state of a $10\times 10$ frustrated quantum magnet—a task that can be formulated as imaginary-time evolution—have already involved approximately $10^6$ variational parameters~\cite{Chen2024Empowering}. In contrast, the number of parameters in quantum circuits, using Trotterization as an example, scales only linearly with evolution time. Because NOVQS is based on the same variational principle as tVMC, they can serve as a variational benchmarking platform for dynamical simulations. Such a platform could guide the design of PQC ansatz and facilitate systematic comparisons of the expressive power and parameter efficiency of PQCs with those of TNS and NQS.

	\section{Acknowledgments}
	The molecular Hamiltonians are generated using PySCF~\cite{pyscf} and OpenFermion~\cite{OpenFermion}. The quantum computing is emulated using PennyLane~\cite{PennyLane} with JAX~\cite{jax} as the interface. 
    The quantum circuit diagrams are generated with the Quantikz LaTeX package~\cite{kay2018tutorial}. 
	The code and data that support the findings of this article are openly available at GitHub~\cite{code}.
	This work is supported by 
	Quantum Science and Technology-National Science and Technology Major Project (2023ZD0300200), 
	the National Natural Science Foundation of China Grant (No.~12361161602), 
	NSAF (Grant No.~U2330201), 
	Beijing Natural Science Foundation Z250004, 
	Beijing Science and Technology Planning Project (Grant No.~Z25110100810000),
	and the High-performance Computing Platform of Peking University.

	\appendix
	
	\section{Single-ancilla quantum phase estimation}\label{app:QPE}
	Beyond the standard quantum phase estimation (QPE) based on the quantum Fourier transform, there is an alternative method known as single-ancilla QPE~\cite{Miroslav2007Arbitrary,Kimmel2015Robust,Lin2022Heisenberg,Jakob2026Phase}, which relies on the Hadamard test and classical signal processing. 
	Our ansatz, expressed as a linear combination of unitaries (LCU) acting on a reference state, can be incorporated into this scheme to estimate the ground-state energy.
	
\begin{figure}[htbp]
		\centering
		\begin{quantikz}
			\lstick{$\ket{0}$} & \gate{H} & \ctrl{1} & \gate{R_{\rm Z}(b_j)} & \gate{H} & \meter{} \\
			\lstick{$\ket{\Psi_0}$} & \qw & \gate{U_j} & \qw & \qw & \qw
		\end{quantikz}
		\caption{The Hadamard test circuit used for single-ancilla QPE. This circuit estimates $\Re( e^{i b_j}\bra{\Psi_0} U_j \ket{\Psi_0})$ via measuring the ancilla qubit in the computational basis. To estimate $\Im (e^{i b_j}\bra{\Psi_0} U_j \ket{\Psi_0})$, just change the rotation angle in the $R_{\rm Z}$ gate from $b_j$ to $b_j - \pi/2$.}
		\label{fig:single-anciila_QPE}
	\end{figure}
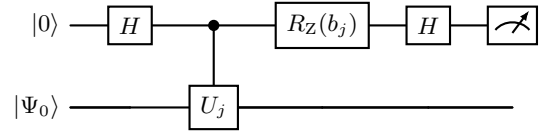

	Within the NOVQS framework, the RTE operator $e^{-iHt}$ is realized as $\lambda \sum_{j=1}^n p_j e^{i b_j} U_j$, where $p_j = a_j / (\sum_{k=1}^n a_k)$ represents the sampling probability of the $j$-th PQC. The sampled unitary $U_j$ and its associated phase $b_j$ are applied to the Hadamard test circuit shown in Fig.~\ref{fig:single-anciila_QPE}. By measuring the ancilla qubit in the computational basis as $\mu = \pm 1$, the time signal $g(t) = \bra{\Psi_0} e^{-iHt} \ket{\Psi_0}$ can be estimated as $\lambda \mathbb{E}[\mu^R + i \mu^I]$, where $\mu^R$ and $\mu^I$ denote the real and imaginary components, respectively. Subsequently, given $g(t_i)$ evaluated at a discrete set of time points $\{t_i\}$, signal processing techniques can extract the ground-state energy with Heisenberg scaling~\cite{Lin2022Heisenberg}, provided that the initial state $\ket{\Psi_0}$ has a non-negligible overlap with the true ground state.
	
	\section{Direct preparation of the wavefunction}\label{app:preparation}
	
	Once the optimized parameters $\{c, \theta\}$ are obtained at discrete time points, we can measure certain properties of the quantum system using the wavefunction defined in Eq.~(\ref{eq:ansatz_normalized}). 
	The fidelity with a goal state, the energy of the system, the entanglement entropy between subsystems, two-point correlations and other observables are all examples to which our method can be straightforwardly applied.
	However, the state in Eq.~(\ref{eq:ansatz}) is described by a linear combination of parameterized quantum states. Its direct realization within a single quantum circuit is required in some cases.
	For example, in QPE, one need to deterministically implement the real time evolution at different time. 
	In what follows, we show that we can construct the LCU~\cite{Childs2012hamiltonian} in Eq.~(\ref{eq:ansatz_normalized}) by two steps~\cite{Berry2015simulating}: block encoding\cite{Low2019hamiltonian} and oblivious amplitude amplification~\cite{Berry2014exponential}.
	
	\subsection{Block encoding}
	The total ansatz can be expressed as a sum of unitaries acting on the inital state: $\ket{\Phi(\theta(t))} = A\ket{\Psi_0}$, where
	\begin{align}
		A = \sum_{j=1}^{N_c} c_j(t) U_j(\theta_j(t)).
	\end{align}
	Our goal is to realize the normalized state $\ket{\tilde{\Phi}(\theta(t))}$ on a quantum circuit.
	We define the block encoding unitary as 
	\begin{align}
		V = \sum_{j=1}^{N_c} \ket{j}_a\bra{j}_a \otimes e^{ib_j} U_j(\theta_j),
	\end{align}
	where the unitary $e^{ib_j} U_j(\theta_j)$ is controlled by the ancilla qubits in the state $\ket{j}_a$. Here, we absorb the phase factors into the controlled unitaries. When the ancillary state is in $\ket{j}_a$, we apply a phase of $e^{i b_j}$. In fact, this forms a diagonal unitary operator $U_{\rm diag} = \text{diag}(e^{i b_1}, \dots, e^{i b_{N_c}})$ acting on the ancilla qubits. Since $V = U_{\rm diag} \left(\sum_{j=1}^{N_c} \ket{j}_a\bra{j}_a \otimes U_j(\theta_j) \right)$, we can implement $V$ in two steps. Diagonal unitary operators constitute a pivotal building block in both Grover’s algorithm and quantum simulation, and as such, have been extensively investigated~\cite{Schuch2003programmable,Welch2014efficient,Zhang2024depth}. Given the gate set $\{R_Z, \text{CNOT}\}$, such transformations can be implemented using quantum circuits with a depth of $O(N_c)$. Next, we prepare the ancillary state 
	\begin{align}
		B\ket{0}_a = \frac{1}{\sqrt{\lambda}} \sum_{j=1}^{N_c} \sqrt{\frac{a_j}{\norm{\ket{\Phi(\theta}}}} \ket{j}_a,
	\end{align}
	where $\lambda = \sum_j a_j / \norm{\ket{\Phi(\theta(t))}} \geq 1$. Define unitary operator
	\begin{align}
		W = (B^\dag \otimes \mathbbm{1}) V (B \otimes \mathbbm{1}),
	\end{align}
	we have 
	\begin{align}
		W \ket{0}_a \ket{\Psi(0)} = \frac{1}{\lambda} \ket{0}_a \ket{\tilde{\Phi}(\theta(t))} + \sqrt{1-\frac{1}{\lambda^2}}\ket{\perp},
	\end{align}
	where $\ket{\perp}$ represents a state that is orthogonal to $\ket{0}_a$ in the ancillary subspace. By measuring the ancilla qubits in the $\ket{0}_a$ state, we can obtain the normalized state $\ket{\tilde{\Phi}(\theta(t))}$ with success probability $1/\lambda$.
	
	\subsection{Oblivious amplitude amplification}
	The oblivious amplitude amplification (OAA) requires $A/\norm{\ket{\Phi(\theta}}$ being unitary, which is true for the real time case. It is shown that when $A/\norm{\ket{\Phi(\theta}}$ is close to the exact time evolution operator $e^{-i H t}$, the error of OAA is well bounded~\cite{Berry2015simulating}. Therefore, we can further amplify the probability of creating the desired state for the variational simulation of RTE. Define the reflection operator $R = 2\Pi - \mathbbm{1}$, where the projector $\Pi=\ket{0}_a\bra{0}_a \otimes \mathbbm{1}$. Let $\sin{\gamma} = 1/\lambda$, where $\gamma \in (0,\pi/2)$. The OAA constructs
	\begin{align}
		S = -W R W^\dag R,
	\end{align}
	such that
	\begin{align}
		&S^l W \ket{0}_a \ket{\Psi(0)} \notag\\
		=& \sin{((2l+1)\gamma)} \ket{0}_a \ket{\bar{\Phi}(\theta(t))}  + \cos{((2l+1)\gamma)} \ket{\perp}.
	\end{align}
	The unitary operator $S$ can be regard as a $2\gamma$ rotation in the 2-dimension subspace spanned by $\{\ket{0}_a \ket{\tilde{\Phi}(\theta(t))}, \ket{\perp}\}$.

	\bibliography{ref}
	
\end{document}